\documentclass[12pt,preprint]{aastex}

\shorttitle{Chemical abundances in the polar disk of NGC4650A}
\shortauthors{Spavone M. et al.}

\begin{document}


\title{Chemical abundances in the polar disk of NGC4650A: implications
for cold accretion scenario\footnote{Based on the data acquired at the
VLT with FORS2, during the observing runs  $<078.B-0580(A)>$ and
$<079.B0177 (A)>$.}}


 \author{M. Spavone\altaffilmark{1,2}, E. Iodice\altaffilmark{2},
   M. Arnaboldi\altaffilmark{3,4}, O. Gerhard\altaffilmark{5},
   R. Saglia\altaffilmark{5,6} and G. Longo\altaffilmark{1,2}}
 \email{spavone@na.infn.it} \affil{Dipartimento di Scienze Fisiche,
   Universit\'a Federico II, via Cinthia 6, I-80126 Napoli, Italy}
 \affil{INAF-Astronomical Observatory of Naples, via Moiariello 16,
   I-80131 Napoli, Italy} \affil{European Southern Observatory,
   Karl-Schwarzschild-Stra$\ss$e 2, D-85748 Garching bei M\"{u}nchen,
   Germany} \affil{INAF, Osservatorio Astronomico di Pino Torinese,
   I-10025 Pino Torinese, Italy} \affil{Max-Plank-Institut f\"{u}r
   Extraterrestrische Physik, Giessenbachstra$\ss$e, D-85741 Garching
   bei M\"{u}nchen, Germany\\ Universit\"{a}tssternwarte M\"{u}nchen,
   Scheinerstra$\ss$e 1, D-81679 M\"{u}nchen, Germany}


%\altaffiltext{1}{PhD student of the Astronomical Observatory of
%Capodimonte, INAF-OAC and University of Naples ''Federico II''.}
%\altaffiltext{2}{Society of Fellows, Harvard University.}
%\altaffiltext{3}{present address: Center for Astrophysics, 60 Garden
%Street, Cambridge, MA 02138} \altaffiltext{4}{Visiting Programmer,
%Space Telescope Science Institute} \altaffiltext{5}{Patron, Alonso's
%Bar and Grill}


\begin{abstract}

The aim of the present study is to test whether the cold accretion of
gas through a "cosmic filament" \citep{Mac06} is a possible formation
scenario for the polar disk galaxy NGC 4650A. If polar disks form from
cold accretion of gas, the abundances of the HII regions may be
similar to those of very late-type spiral galaxies, regardless of the
presence of a bright central stellar spheroid, with total luminosity
of few $10^{9} L_{\odot}$.

We use deep long slit spectra obtained with the FORS2 spectrograph at
the VLT in the optical and near-infrared wavelength ranges for the
brightest HII regions in the disk polar disk of NGC 4650A. The
strongest emission lines ([OII] $H_{\beta}$, [OIII], $H_{\alpha}$) were used
to derived oxygen abundances, metallicities and the global star
formation rates for the disk.  The deep spectra available allowed us
to measure the Oxygen abundances ($12 + log (O/H)$) using the {\it
  Empirical method} based on intensities of the strongest emission
lines, and the {\it Direct method}, based on the determination of the
electron temperature from the detection of weak auroral lines, as the
[OIII] at 4363 \AA.

The Oxygen abundance measured for the polar disk is then compared with
those measured for different galaxy types of similar total
luminosities, and then compared against the predictions of different
polar ring formation scenarios.

The average metallicity values for the polar disk in NGC 4650A is
$Z=0.2 Z_{\odot}$, and it is lower that the values measured for
ordinary spirals of similar luminosity. Moreover the gradient of the
metallicity is flat along the polar disk major axis, which implies
none or negligible metal enrichment from the stars in the older
central spheroid.

The low metallicity value in the polar disk NGC 4650A and the flat
metallicity gradient are both consistent with a later infall of
metal-poor gas, as expected in the cold accretion processes.

\end{abstract}





\section{Introduction} \label{intro}

The hierarchical, merger-dominated picture of galaxy formation is
based on the Cold Dark Matter (CDM) model \citep{Col00}, which
predicts that the observed galaxies and their dark halo (DH) were
formed through a repeated merging process of small systems
(\citealt{Del06}; \citealt{Gen08}). In this framework, major and minor
mergers of disky systems do play a major role in the formation of
spheroid and elliptical galaxies (\citealt{Naa07}; \citealt{Bou07}),
in all environments and from the Local Group to high-redshift universe
\citep{Con03}. The gas fraction is a key parameter in the physics of
such gravitational interactions: if it is high enough, an extended and
massive disk structure can survive (\citealt{Spr05}; \citealt{Rob08}).
Galaxies can get their gas through several interacting processes,
such as smooth accretion, stripping and accretion of primordial
gas, which are equally important in the growth of galaxies.  Recent
theoretical works have argued that the accretion of external gas from
the cosmic web filaments, with inclined angular momentum
(\citealt{Dav01}, \citealt{Sem05}), might be the most realistic way by
which galaxies get their gas. This process may also explain the
build-up of high redshift disk galaxies (\citealt{Ker05},
\citealt{Ker08}, \citealt{Bro08}, \citealt{Dek09},
\citealt{Bou09}). The relative share of all gravitational
interactions depends on the environments and it drives many
morphological features observed in galaxies, such as bars and polar
rings.

Galaxies with polar rings (PRGs) generally contain a central
featureless stellar spheroid and an elongated structure, the ``polar
ring'' (made up by gas, stars and dust), which orbits in a nearly
perpendicular plane to the equatorial one of the central galaxy
\citep{Whi90}. The decoupling of the angular momentum of the polar
structure and the central spheroid cannot be explained by the collapse
of a single protogalactic cloud: thus a ``second event'' must have
happened in the formation history of these systems. This is the reason
why studying PRGs promise to yield detailed information about many of
the processes at work during galaxy interactions and merging
(\citealt{Iod02}, \citealt{Iod02A}, \citealt{Res94}, \citealt{Res02},
\citealt{Bou03}).

The debate on the origin of PRGs is still open and two main
  processes have been proposed {\it i)} a major dissipative merger or
  {\it ii)} gas accretion.  In the merging scenario, the PRG results
  from a ``polar'' merger of two disk galaxies with unequal mass,
  (\citealt{Bek97}; \citealt{Bek98}; \citealt{Bou05}): the morphology
  of the merger remnants depends on the merging initial orbital
  parameters and the initial mass ratio of the two galaxies.  In the
  accretion scenario, the polar ring may form by a) the disruption of
  a dwarf companion galaxy orbitating around an early-type system, or
  by b) the tidal accretion of gas stripping from a disk galaxy
  outskirts, captured by an early-type galaxy on a parabolic encounter
  (\citealt{Res97}; \citealt{Bou03}; \citealt{Han09}). In the latter
  case, the total amount of accreted gas by the early-type object is
  about $10\%$ of the gas in the disk donor galaxy, i.e. up to
  $10^{9}$ $M_{\odot}$. Both major merger and accretion scenarios are
  able to account for many observed PRGs morphologies and kinematics,
  such as the existence of both wide and narrow rings, helical rings
  and double rings \citep{Whi90}.

Very recently, a new mechanism has been proposed for the formation of
wide disk-like polar rings: a long-lived polar structure may form
through cold gas accretion along a filament, extended for $\sim 1$
Mpc, into the virialized dark matter halo \citep{Mac06}. In this
formation scenario, there is no limits to the mass of the accreted
material, thus a very massive polar disk may develop either around a
stellar disk or a spheroid.  \citet{Bro08}, by using high-resolution
cosmological simulations of galaxy formation, have confirmed and
strengthened the formation scenario proposed by \citet{Mac06}.  In
this case polar disk galaxies can be considered as extreme examples of
angular momentum misalignment that occurs during the hierarchical
structure formation. In the merging history, an inner disk formed
first after the last major merger of two galaxies with a 1:1 mass
ratio, then, due to the high gas fraction, the galaxy rapidly forms a
new disk whose angular momentum depends on the merger orbital
parameters. At later times, gas continues to be accreted along the
disk which may be also in a plane perpendicular to the inner disk. The
morphology and kinematics of one simulated object, in both
simulations, are similar to those observed for NGC4650A: in
particular, \citet{Bro08} found that such formation mechanism can
self-consistently explain both morphology and kinematics of central
spheroid and polar structure, and all the observed features (like
colors and colors gradient, longevity, spiral arms, HI content and
distribution).

\begin{figure*}
\centering
\includegraphics[width=7cm]{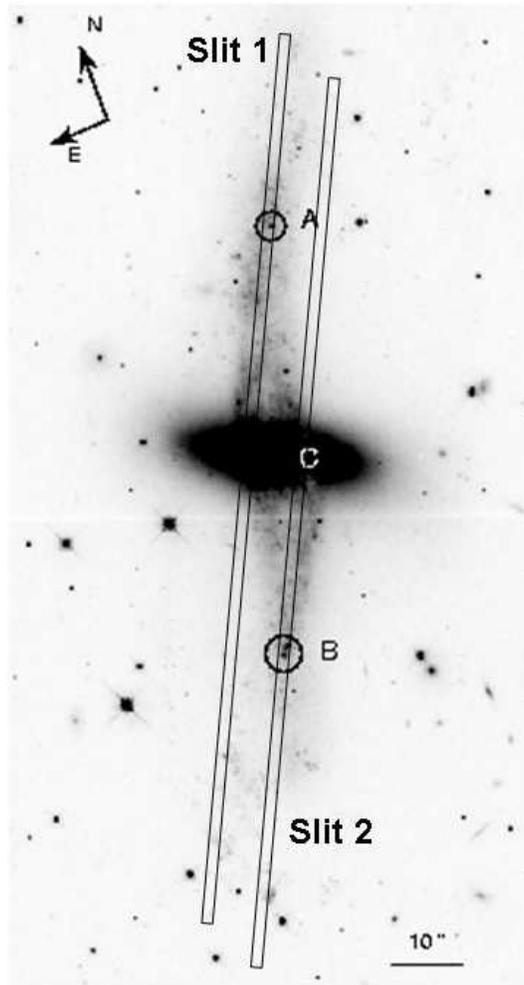}
\caption{Optical image of NGC4650A with superimposed the slits used to
  acquire data analyzed in this work.} \label{slit}
\end{figure*}

NGC~4650A is the prototype for PRGs (see Fig. \ref{slit}). Its
luminous components, inner spheroid and polar structure were
studied in optical and near-infrared (NIR) photometry,
spectroscopy, and in the radio emission, HI 21 cm line and continuum
(\citealt{Arn97}, \citealt{Gal02}, \citealt{Iod02}, \citealt{Swa03},
\citealt{Iod06}).

 The polar structure in NGC~4650A is a disk, very similar to that
   of late-type spirals or LSB galaxies, rather than a ring. The polar
   disk stars and dust can be reliably traced to $\sim 1.2$ kpc radius
   from the galaxy nucleus, and the surface brightness profiles have
   an exponential decrease (\citealt{Iod02},
   \citealt{Gal02}). Furthermore, the rotation curves measured from
 the emission and absorption line optical spectra are consistent with
 those of a disk in a differential rotation rather than that of a
 narrow ring \citep{Swa03}. This is also confirmed by the HI 21 cm
   observations \citep{Arn97} which show that the gas is five times
 more extended than the luminous polar structure, with a
 position-velocity diagram very similar to those observed for edge-on
 disks. The polar disk is very massive, since the total HI mass is
 about $10^{10} M_{\odot}$, which added to the mass of stars, makes
 the mass in the polar disk comparable with the total mass in the
 central spheroid \citep{Iod02}. The morphology of the central
   spheroid resembles that of a low-luminosity early-type galaxy: the
   surface brightness profile is described by an exponential law, with
   a small exponential nucleus; its integrated optical vs. NIR colors
   are similar to those of an intermediate age stellar population
   (\citealt{Iod02}, \citealt{Gal02}).  New high resolution
 spectroscopy in NIR on its photometric axes suggests that this
 component is a nearly-exponential oblate spheroid supported by
 rotation \citep{Iod06}.

These new kinematic data, together with the previous studied, set
  important constraints on the possible formation mechanisms for
  NGC4650A.

Because of the extended and massive polar disk, with strong $H_\alpha$
emissions, NGC 4650A is the ideal candidate to measure the chemical
abundances, metallicity and star formation rates (SFR) via
spectroscopic measurements of line emissions along the major axis of
the polar disk. The goal is to compare the derived values for the
metallicity and SFR with those predicted by the different formation
scenarios.  As we shall detail in the following sections, if the polar
structure forms by accretion of primordial cold gas from cosmic web
filament, we expect the disk to have lower metallicities of the order
of $Z\sim 1/10 Z_{\odot}$ \citep{Age09} with respect to those of
same-luminosity spiral disks.

We shall adopt for NGC4650A a distance of about 38 Mpc based on $H_{0}
= 75 \ km \ s^{-1} \ Mpc^{-1}$ and an heliocentric radial velocity $V
= 2880\ km \ s^{-1}$, which implies that 1 arcsec = 0.18 kpc.

\section{Observations and data reduction}\label{obs}

%% In a manner similar to \objectname authors can provide links to dataset
%% hosted at participating data centers via the \dataset{} command.  The
%% second curly bracket argument is printed in the text while the first
%% parentheses argument serves as the valid data set identifier.  Large
%% lists of data set are best provided in a table (see Table 3 for an example).
%% Valid data set identifiers should be obtained from the data center that
%% is currently hosting the data.
Spectra were obtained with FORS2@UT1, on the ESO VLT, in service
mode, during two observing runs: 078.B-0580 (on January 2007) and
079.B-0177 (on April 2007). FORS2 was equipped with the MIT
CCD~910, with an angular resolution of $0.25''$ pixel$^{-1}$. The
adopted slit is $1.6 ''$ wide and $6.8'$ long. Spectra were
acquired along the North and South side of the polar disk, at
$P.A.=152^{\circ}$ (see Fig. \ref{slit}), in order to include the
most luminous HII regions in polar disk. The total integration
time for each direction is 3 hours, during the 078.B-0580 run and
2.27 hours during 079.B-0177 run, respectively, with an average
seeing of $1.2 ''$.

At the systemic velocity of NGC~4650A, to cover the red-shifted
emission lines of the $[OII]\lambda3727$, $[H_{\gamma}]\lambda4340$,
$[OIII]\lambda4363$, $[OIII]\lambda\lambda4959,5007$,
$[H_{\beta}]\lambda4861$, $[NII]\lambda5755$ the grism GRIS-600B+22
was used in the $3300-6210$ \AA\ wavelength range, with a dispersion
of 50 \AA/mm (0.75 \AA/pix). In the near-infrared
$5600-11000$\AA\ wavelength range, the grism GRIS-200I+28 was used,
with a dispersion of 162 \AA/mm (2.43 \AA/pix) to detect the fainter
$[SII]\lambda\lambda6717,6731$ and $[SIII]\lambda\lambda9068,9532$
emission lines, with a $S/N \geq 20$ and the brighter
$[H_{\alpha}]\lambda6563$ emission line, with a $S/N >150$. In this
wavelength range, where the sky background is much more variable and
higher with respect to the optical domain, to avoid the saturation of
the sky lines a larger number of scientific frames with short
integration time (850 sec) were acquired.

The data reduction was carried out using the {\small CCDRED} package
in the IRAF\footnote{IRAF is distributed by the National Optical
  Astronomy Observatories, which is operated by the Associated
  Universities for Research in Astronomy, Inc. under cooperative
  agreement with the National Science Foundation.} ({\it Image
  Reduction and Analysis Facility}) environment. The main strategy
adopted for each data-set included dark subtraction\footnote{Bias
  frame is included in the Dark frame.}, flat-fielding correction, sky
subtraction and rejection of bad pixels. Wavelength calibration was
achieved by means of comparison spectra of Ne-Ar lamps acquired for
each observing night, using the IRAF TWODSPEC.LONGSLIT package. The
sky spectrum was extracted at the outer edges of the slit, for $r \ge
40$ arcsec from the galaxy center, where the surface brightness is
fainter than $24 mag/arcsec^2$, and subtracted off each row of the two
dimensional spectra by using the IRAF task BACKGROUND in the
TWODSPEC.LONGSLIT package.  On average, a sky subtraction better than
$1\%$ was achieved. The sky-subtracted frames, both for North and
South part of the polar disk, were co-added to a final median averaged
2D spectrum.

\begin{figure*}
\centering
\includegraphics[width=10cm]{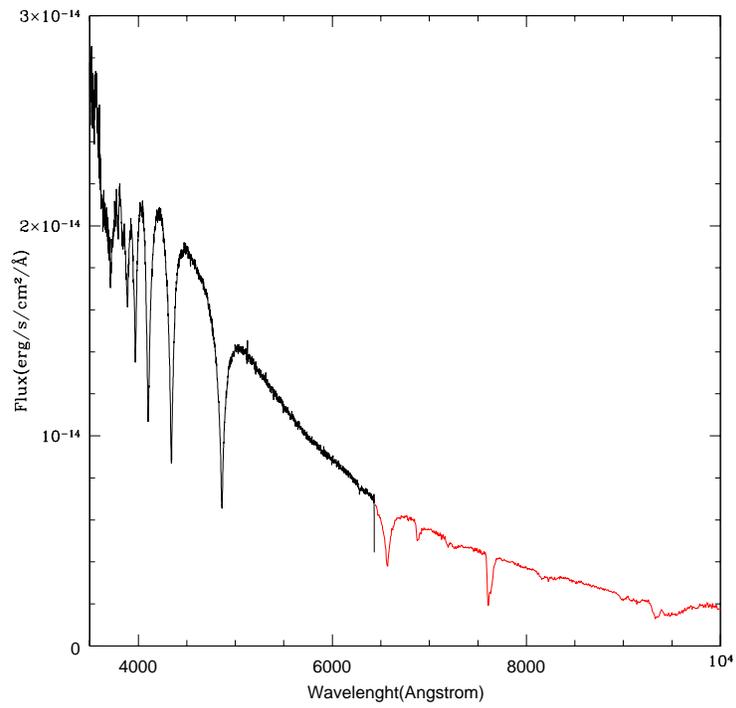}
\caption{1-D spectrum of the standard star LTT4816, used to
flux-calibrate the spectra. The black line represent the spectrum of
the star acquired in the blue wavelength range, while the red line is
the same star acquired in the red range.} \label{std}
\end{figure*}

The final step of the data-processing is the flux calibration of each
2D spectra, by using observations of the standard star LTT4816 and the
standard tasks in IRAF (STANDARD, SENSFUNC and CALIBRATE).  The flux
calibration is very important in the present study because we need to
``align'' two spectra, which cover different wavelength range and
taken in different times. Thus, we checked and obtained that the
calibrations for the spectra in both spectral range were
consistent. Fig. \ref{std} shows the 1D flux calibrated spectra of the
spectrophotometric standard star used to calibrate the spectra in the
whole range $3300-11000$ \AA. To perform the flux calibration we
extracted a 1-D spectrum of the standard star to find the calibration
function; then we extracted a set of 1-D spectra of the galaxy summing
up a number of lines corresponding to the slit width. Since the slit
width was $1.3 ''$ and the scale of the instrument was $0.25 ''/pix$,
we collapsed seven lines to obtain each 1-D spectrum. Finally we
applied the flux calibration to this collection of spectra.

Furthermore, we compared our flux calibrated spectra with others
acquired at the Siding Spring Observatory with the Double Beam
Spectrograph (DBS) \citep{But06}.  The DBS has a dichroic that slits
the light in a red and a blue arm, therefore the flux calibration with
standard stars can be done simultaneously for the red and the blue
arms. We used these spectra to check for any difference in the flux
calibrations, finding that our flux calibrated spectra, both of the
template star and of the galaxy, turn out to be consistent with them.

\begin{figure*}
  \centering
      \includegraphics[width=12cm]{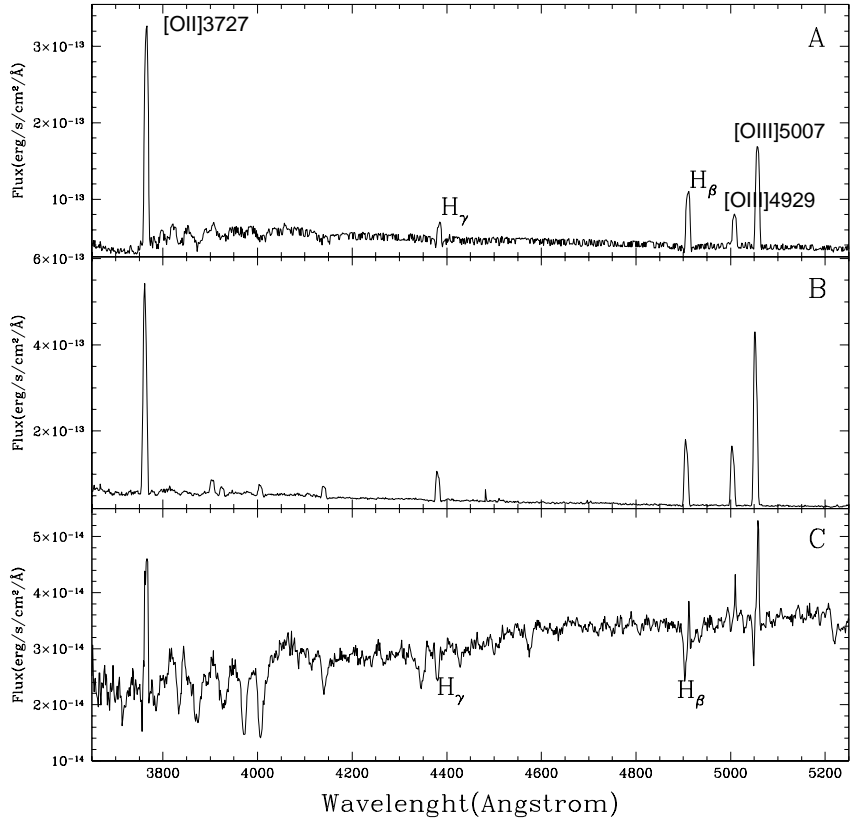}
\caption{\emph{Top panel} - Spectrum of NGC4650A in the blue
  wavelength range, corresponding to the region marked as A in the
  Fig. \ref{slit}. \emph{Middle panel} - Spectrum of the region B in
  the Fig. \ref{slit}. \emph{Bottom panel} - Spectrum of the region C
  in the Fig. \ref{slit}, where are clearly visible the absorbing
  features of the line $H_{\gamma}$ and $H_{\beta}$.}
       \label{spec_blu}
\end{figure*}

\begin{figure*}
  \centering
      \includegraphics[width=12cm]{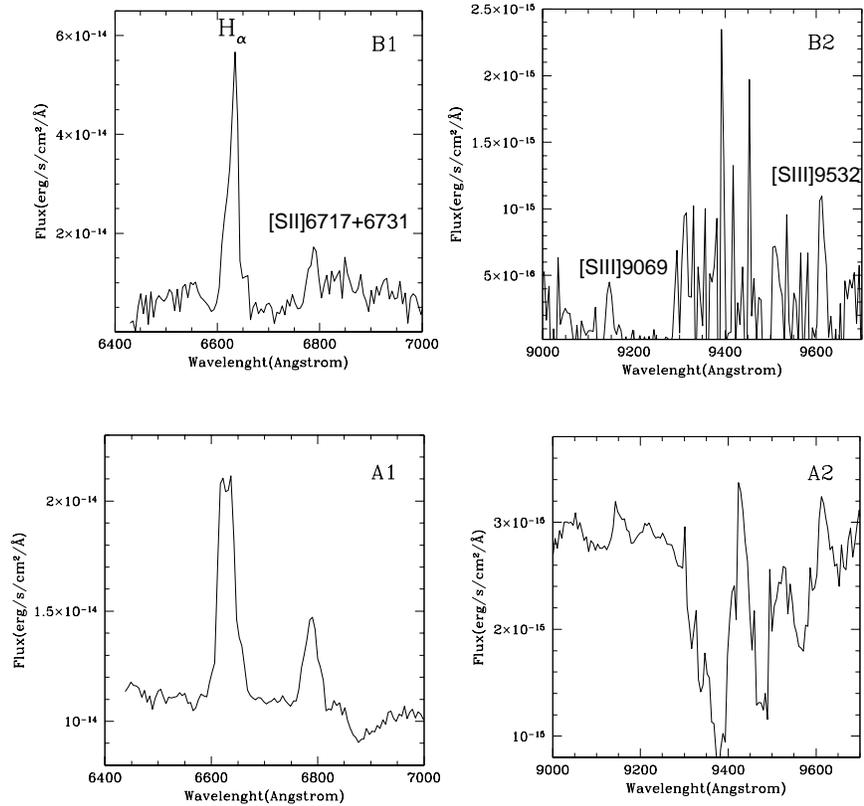}
\caption{\emph{Top panel} - Spectra of NGC4650A in the red wavelength
  range, corresponding to the region marked as A in the
  Fig. \ref{slit}. We divided the spectra in A1 and A2 in order to
  obtain a better visualization of the lines. \emph{Bottom panel} -
  Spectrum of the region B (B1 and B2 as explained below) in the
  Fig. \ref{slit}.}
       \label{spec_red}
\end{figure*}

The wavelength and flux-calibrated spectra are shown in
Fig. \ref{spec_blu} and \ref{spec_red}. In the blue spectrum (top
panel) are clearly visible a number of emission lines: $H_{\beta}$,
$H_{\gamma}$, $[OII]\lambda3727$ and $[OIII]\lambda\lambda4959,5007$,
while in the red one (bottom panel) we have $H_{\alpha}$ (blended with
the $[NII]\lambda6583$ line appearing in the red wing of
$H_{\alpha}$), $[SII]\lambda\lambda6717,6731$ and
$[SIII]\lambda\lambda9069,9532$. From a two-Gaussian fit to the
combined emission, we estimate the line ratio $[NII]\lambda6583
/(H_{\alpha}+[NII]\lambda6548 \simeq 0.1$.  For this reason the
$H_{\alpha}$ flux is that measured as the total flux in the line
reduced by the contribution of $[NII]\lambda6583$. The observed
emission lines and their uncorrected and reddening-corrected fluxes
relative to $H_{\beta}$ are listed in Tab. \ref{fluxopt} and
Tab. \ref{fluxnir}.

Since ground-based near-infrared spectroscopy is affected by the
strong and variable absorption features due to the Earth's atmosphere,
we accounted for this effect in our spectra, in which the telluric
absorption bands are clearly visible around 7200, 8200 and 9300
\AA. In order to perform this correction we used the ''telluric
standard star'' LTT4816, near both in time and air mass to the object,
and the IRAF task TELLURIC; by fitting the continuum, telluric
calibration spectra are shifted and scaled to best divide out telluric
features from data spectra. The ratios of the uncorrected fluxes
relative to those obtained applying the average telluric correction
are the following: $[SII]\lambda 6717_{uncorrected}/ [SII]\lambda
6717_{corrected} = [SII]\lambda 6731_{uncorrected}/ [SII]\lambda
6731_{corrected} = 1.7$ and $[SIII]\lambda 9069_{uncorrected}/
[SIII]\lambda 9069_{corrected} = [SIII]\lambda 9532_{uncorrected}/
[SIII]\lambda 9532_{corrected} = 0.6$.

\subsection{Measurement of emission-lines fluxes}

The fluxes of the above mentioned emission lines were measured using
the IRAF {\small SPLOT} routine, that provides an interactive facility
to display and analyze spectra. The $H_{\beta}$ is evaluated for $r
\geq 10$ arcsec, where only the emission line is present; for lower
distances, i.e. where stars relative to the spheroid also contributes
to the spectra, the $H_{\beta}$ is also in absorption. We evaluated
flux and equivalent width by marking two continuum points around the
line to be measured.  The linear continuum is subtracted and the flux
is determined by simply integrating the line intensity over the local
fitted continuum. The errors on these quantities have been calculated,
following \citet{Per03}, by the relation $\sigma_{1} =
\sigma_{c}N^{1/2}[1+EW/(N\Delta)]^{1/2}$, were $\sigma_{1}$ is the
error in the line flux, $\sigma_{c}$ is the standard deviation in a
box near the measured line and represents the error in the continuum
definition, N is the number of pixels used to measure the flux, EW is
the equivalent width of the line and $\Delta$ is the wavelength
dispersion in \AA/pixel. The errors relative to each emission line
fluxes are listed in Tab. \ref{fluxopt} and Tab. \ref{fluxnir}.

%% In this section, we use  the \subsection command to set off
%% a subsection.  \footnote is used to insert a footnote to the text.


%% This section also includes several of the displayed math environments
%% mentioned in the Author Guide.

\subsection{Reddening correction}\label{reddening}

Reduced and flux calibrated spectra and the measured emission line
intensities were corrected for the reddening, which account both for
that intrinsic to the source and to the Milky Way. By comparing the
intrinsic Balmer decrements, $H_{\alpha}/H_{\beta}=2.89$ and
$H_{\gamma}/H_{\beta}=0.468$, predicted for large optical depth (case
B) and a temperature of $10^4$ K, with the observed one, we derived
the visual extinction $A(V)$ and the color excess $E(B-V)$, by
adopting the mean extinction curve by \citet{Card89} $A(\lambda)/A(V)
= a(x)+b(x)R_{V}$, where $R_{V}[\equiv\ A(V)/E(B-V)]=3.1$ and
$x=1/\lambda$.
In order to estimate reddening correction for all the observed
emission lines, we used the complete extinction
curve into three wavelengths regions (\citet{Card89}):
\emph{infrared} ($\lambda\geq0.9 \mu m$), \emph{optical/NIR}
($0.9\mu m\geq\lambda\geq0.3 \mu m$) and \emph{ultraviolet} ($0.125\mu
m\geq\lambda\geq0.10 \mu m$), which are characterized by different
relations of a(x) and b(x). All the emission lines in our spectra are
in the \emph{optical/NIR} range, except for the $[SIII]\lambda9532$,
that falls in the \emph{infrared} range, so we used the average
$R_{V}$-dependent extinction law derived for these intervals to
perform the reddening correction.\\ We measured the fluxes of
$H_{\beta}$, $H_{\gamma}$ and $H_{\alpha}$ lines at each distance from
the galaxy center and for each spectra (North slit and South slit),
than we derived the average observed Balmer decrements, which are the
following:\\ $H_{\alpha}/H_{\beta}$ = $2.40 \pm\ 0.01$\\
$H_{\gamma}/H_{\beta}$ = $0.41 \pm\ 0.01$\\ while the color excess
obtained by using these observed decrements are:\\
$[E(B-V)]_{H_{\alpha}/H_{\beta}}$ = $0.20 \pm\ 0.004$\\
$[E(B-V)]_{H_{\gamma}/H_{\beta}}$ = $0.25 \pm\ 0.012$.\\ Such values
of E(B-V) are used to derive the extinction $A_{\lambda}$, through the
Cardelli's law; in particular, the $[E(B-V)]_{H_{\gamma}/H_{\beta}}$
and $[E(B-V)]_{H_{\alpha}/H_{\beta}}$ are used respectively for the
reddening correction in optical and NIR wavelength range.  The
corrected fluxes are given by $F^{\lambda}_{int} / F^{H_{\beta}}_{int}
= F^{\lambda}_{obs} / F^{H_{\beta}}_{obs}
10^{0.4[A_{\lambda}-A_{H_{\beta}}]}$. Observed and reddening-corrected
emission line fluxes are reported in Tab. \ref{fluxopt} and
Tab. \ref{fluxnir}.

%% The equation environment wil produce a numbered display equation.

%% The \notetoeditor{TEXT} command allows the author to communicate
%% information to the copy editor.  This information will appear as a
%% footnote on the printed copy for the manuscript style file.  Nothing will
%% appear on the printed copy if the preprint or
%% preprint2 style files are used.

%% The eqnarray environment produces multi-line display math. The end of
%% each line is marked with a \\. Lines will be numbered unless the \\
%% is preceded by a \nonumber command.
%% Alignment points are marked by ampersands (&). There should be two
%% ampersands (&) per line.

%% Putting eqnarrays or equations inside the mathletters environment groups
%% the enclosed equations by letter. For instance, the eqnarray below, instead
%% of being numbered, say, (4) and (5), would be numbered (4a) and (4b).
%% LaTeX the paper and look at the output to see the results.

\section{Oxygen abundances determination}\label{met}

%The analysis of nebular spectra in HII regions is the best tool for
%the determination of chemical abundances in spiral and irregular
%galaxies. However this is not a simple matter, in fact it is well
%known that the ionic abundances depend on the intensity of the
%emission lines involved and on the electronic density and
%temperature. The abundances of several elements can in principle be
%determined by using strong emission lines clearly visible in the
%spectra, but the physical conditions in the HII regions should be
%taken into account, via the determination of the electron
%temperature. This can be obtained by measuring appropriate line
%ratios, that unfortunately involve intrinsically weak lines which can
%be detected and measured only for the brighter and hotter objects.

The main aim of the present work is to derive the chemical abundances
in the polar disk of NGC4650A: in what follow we evaluate the oxygen
abundances following the methods outlined in \citet{Pag79},
\citet{Diaz00} and \citet{Pil01}, referring to them as \emph{Empirical
  methods}, and those introduced by \citet{Ost89} and \citet{All84},
or \emph{Direct methods}.

The large dataset available (presented
in Sec. \ref{obs}) let us to investigate both the \emph{Empirical
methods}, based on the intensities of easily observable lines
(Sec. \ref{Emp}), and the \emph{Direct method}, based on the
determination of the electron temperature, by measuring the intensities
of the weak auroral lines (Sec. \ref{dir}).\\ As described in details
in the next sections, we have derived the oxygen abundance parameter
$12+log(O/H)$ along the polar disk, by using both the
\emph{Empirical} and \emph{Direct} methods.

\subsection{Empirical oxygen and sulphur abundance
measurements}\label{Emp}

The \emph{Empirical methods} are based on the cooling properties
of ionized nebulae which translate into a relation between
emission-line intensities and oxygen abundance. Several abundance
calibrators have been proposed based on different emission-line
ratios: among the other, in this work we used $R_{23}$
\citep{Pag79}, $S_{23}$ \citep{Diaz00} and the P-method
\citep{Pil01}. The advantages of different calibrators have been
discussed by several authors (\citealt{Diaz00}, \citealt{Kob99},
\citealt{Kob04}, \citealt{Per05}, \citealt{Kew08}, \citealt{Hid09}).

The method proposed by \citet{Pag79} is based on the variation of
the strong oxygen lines with the oxygen abundance. \citet{Pag79}
defined the ''oxygen abundance parameter'' $R_{23} = ([OII]\lambda
3727 + [OIII]\lambda \lambda 4959 + 5007)/H_{\beta}$, which
increase with the oxygen abundance for values lower than 20\% of
the solar one and then reverses his behavior, decreasing with
increasing abundance, due to the efficiency of oxygen as a cooling
agent that lead to a decreasing in the strength of the oxygen
emission lines at higher metallicities. Three different regions
can be identified in the trend of $R_{23}$ with the oxygen
abundance \citep{Per05}: a lower branch ($12 + log(O/H) < 8.1$),
in which $R_{23}$ increase with increasing abundance, an upper
branch ($12 + log(O/H) > 8.4$), in which the trend is opposite,
and a turnover region ($8.1 < 12 + log(O/H) < 8.4$). While the
upper and lower branches can be fitted by regression lines, the
turnover region seems to be the most troublesome, in fact object
with the same value of $R_{23}$ can have very different oxygen
abundances. The $R_{23}$ parameter is affected by two main
problems: the two-valued nature of the calibration and the
dependence on the degree of ionization of the nebula.

To break the degeneracy that affects the metallicity for values of
$12 + log(O/H) \geq 8.0$, \citet{Diaz00} proposed an alternative
calibrator, based on the intensity of sulphur lines: $S23=
([SII]{\lambda}{\lambda} 6717 + 6731 + [SIII]{\lambda}{\lambda}
9069 + 9532)/H_{\beta}$. The [SII] and [SIII] lines are analogous
to the optical oxygen lines [OII] and [OIII], but the relation
between the ''Sulphur abundance parameter $S_{23}$'' and the
oxygen abundance remain single valued up to solar metallicities
(\citealt{Diaz00} and \citealt{Per05}). Moreover, because of their
longer wavelength, sulphur lines are less sensitive to the
effective temperature and ionization parameter of the nebula, and
almost independent of reddening, since the [SII] and [SIII] lines
can be measured nearby hydrogen recombination lines.
\citet{Diaz00} have attempted a calibration of oxygen abundance
through the sulphur abundance parameter, obtaining the
following empirical relation:

\begin{equation}\label{diaz}
12 + log(O/H) = 1.53log S_{23}+8.27
\end{equation}

For the polar disk of NGC4650A, by using the emission line fluxes
given in Tab. \ref{fluxopt} and Tab. \ref{fluxnir}, we derived both
$R_{23}$ and $S_{23}$ parameters, which are listed in
Tab. \ref{rs23}. Since there are few measurements available for the
Sulphur lines, we have obtained an average value of $S_{23}$ for the
polar disk, by using the following telluric corrected fluxes ratios: $[SII]\lambda 6717/H_{\beta} = 0.18$,
$[SII]\lambda 6731/H_{\beta} = 0.2$, $[SIII]\lambda 9069/H_{\beta} = 0.43$ and
$[SIII]\lambda 9532/H_{\beta} = 0.25$, obtaining $log S_{23} = 0.024 \pm 0.030$.  The average oxygen
abundance, derived by adopting the Eq. \ref{diaz}, is the following:
$12+log(O/H)_{S} = 8.3 \pm 0.2$.

\begin{figure*}
\centering
\includegraphics[width=11cm]{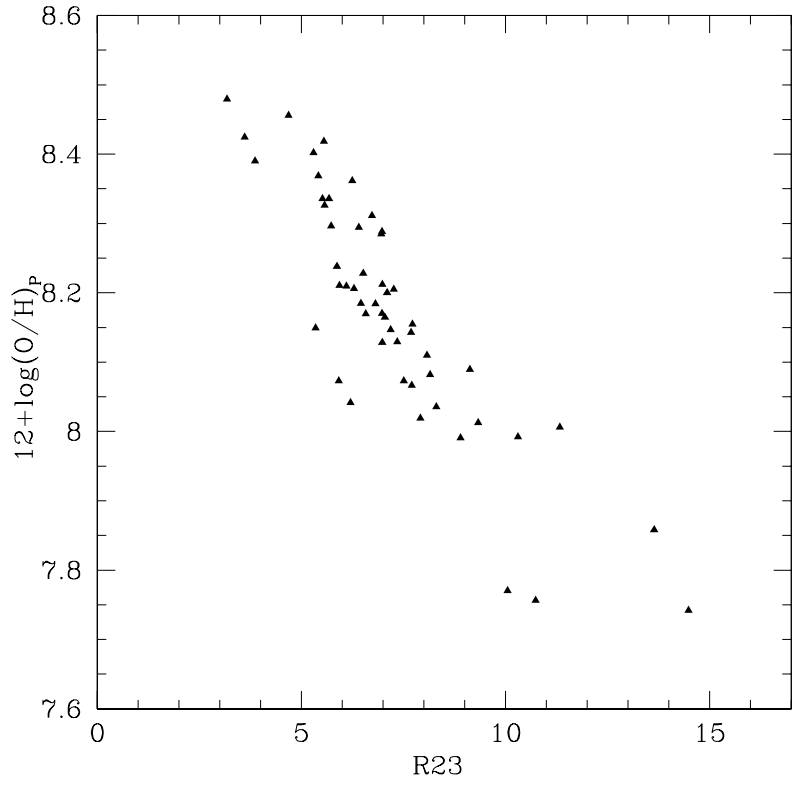}
\caption{Oxygen abundance, obtained by using the empirical calibration introduced by \citet{Pil01},
versus oxygen abundance parameter $R_{23}$ (see Sec. \ref{Emp}).}
\label{Pil}
\end{figure*}

\citet{Pil01} realized that for fixed oxygen abundances the value
of $X_{23} = log R_{23}$ varies with the excitation parameter $P =
R_{3}/R_{23}$, where $R_{3} = OIII[4959+5007]/H_{\beta}$, and proposed
that this latter parameter could be used in the oxygen abundance
determination. This method, called ''P-method'', propose to use a more
general relation of the type $O/H = f(P, R_{23})$, compared with the
relation $O/H = f(R_{23})$ used in the $R_{23}$ method. The equation
related to this method is the following
\begin{equation}\label{pil_cal}
    12+log(O/H)_{P} = \frac{R_{23}+54.2+59.45P+7.31P^{2}}{6.07+6.71P+0.371P^{2}+0.243R_{23}}
\end{equation}
where $P = R_{3}/R_{23}$. It can be used for oxygen abundance
determination in moderately high-metallicity HII regions with
undetectable or weak temperature-sensitive line ratios
\citep{Pil01}.\\ We used also the ''P-method'' to derive the oxygen
abundance in the polar disk of NGC4650A: the values of
$12+log(O/H)_{P}$ at each distance from the galaxy center are listed
in Tab. \ref{rs23}, and shown in Fig. \ref{Pil} and Fig. \ref{fit}. The
average value of the oxygen abundance is $12+log(O/H)_{P} = 8.2 \pm
0.2$, which is consistent with the values derived by using the Sulphur
abundance parameter (Eq. \ref{diaz}).

The metallicities corresponding to each value of oxygen abundances
given before have been estimated. We adopted $12 + log(O/H)_{\odot} =
8.83 = A_{\odot}$ and $Z_{\odot} = 0.02$ \citep{Asp04}. Given that
$Z_{NGC4650A} \approx K Z_{\odot}$ and $K = 10^{[A_{NGC4650A} -
A_{\odot}]}$, we obtain a metallicity for the HII regions of the polar
disk in NGC4650A {\bf $Z \simeq 0.004$ which correspond to $Z \simeq
(0.2 \pm 0.002) Z_{\odot}$}.

\subsection{Direct oxygen abundance measurements}\label{dir}

The electron temperature is the fundamental parameter to directly
derive the chemical abundances in the star-forming regions of
galaxies. In a temperature-sensitive ion, with a well separated
triplet of fine-structure terms, electron temperature and electron
density ($N_e$), can be measured from the relative strengths of lines
coming from excited levels at different energies above the ground
state (see \citealt{Ost89}; \citealt{All84}). As explained in detail
below, the oxygen abundances are function of both the emission line
fluxes and electron temperatures: this relation is the basis of the
commonly known as {\it direct} method or $T_e$ method.\\
Usually, it happens that not all these lines can be observed in the
spectra, or they are affected by large errors. Thus, some
assumption were proposed for the temperature structure through the
star-forming region: for HII galaxies it is assumed a {\it
two-zone model} with a low ionization zone, where the OII, NII,
NeII and SII lines are formed, and a high ionization zone where
the OIII, NIII, NeIII and SIII lines are formed (see, e.g.
\citealt{Cam86}; \citealt{Gar92}). Photoionization models are then
used to relate the temperature of the low ionization zone $t_2$ to
$t_3$, the temperature of the high ionization zone (\citealt{Pag92};
\citealt{Per03}; \citealt{Pil06}; \citealt{Pil07}; \citealt{Pil09}).

For NGC4650A, we aim to derive the oxygen abundance of the polar
disk directly by the estimate of the $O^{++}$ and $O^{+}$ ions
electron temperatures. According to \citet{Izo05} and
\citet{Pil06}, we have adopted the following equations for the
determination of the oxygen abundances, which are based on a
five-level atom approximation:

\begin{equation}\label{OHIII}
12+log(OIII/H) = log([4959+5007]/H_{\beta})+6.2+1.251/t_{3}-0.55logt_{3}-0.014t_{3}
\end{equation}

\begin{equation}\label{OHII}
12+log(OII/H)=log([3727]/H_{\beta})+5.961+1.676/t_{2}-0.40logt_{2}-0.034t_{2}+log(1+1.35x_2)
\end{equation}

where $t_{3}$ and $t_{2}$ are the electron temperatures within the
[OIII] and [OII] zones respectively, in units of $10^4$ K;
$x_{2}=10^{-4}N_{e}t_{2}^{-0.5}$, where $N_{e}$ is the electron
density. The total oxygen abundance is $O/H =OIII/H + OII/H$.

The electron temperatures and the electron density, needed to solve
the above relations, have been derived using the task TENDEM of the
STSDAS package in IRAF, which solves the equations of the statistical
equilibrium within the five-level atom approximation (\citealt{Der87};
\citealt{Sha95}). According to this model, both $t_3$ and $t_2$ are
function of the electron density $N_e$ and of the line ratios
$R_{OIII} = [OIII]\lambda(4959+5007)/[OIII]\lambda4363$ and $R_{OII} =
[OII]\lambda3727/[OII]\lambda7325$, respectively. For NGC4650A, $N_e$
has been derived from the line ratio
$[SII]\lambda6717/[SII]\lambda6731$, which has the average value of
$0.83 \pm 0.06$, through the whole disk extension, for a default
temperature of $10^4$ K. This let to $N_e \sim 1217$ $cm^{-3}$. Even
though the auroral line $[OIII]\lambda4363$ is usually very weak, even
in the metal-poor environments, the large collecting area of the 8m
telescope and the high signal-to-noise spectra, let us to measure such
emission line along the whole extension of the polar disk (see
Tab. \ref{fluxopt}), which let us to estimate $R_{OIII}$.\\ The
electron temperature $t_3=f(R_{OIII},N_e)$ has been calculated by
putting $N_e$ and $R_{OIII}$ as inputs in the TENDEM task: given the
large spread of $R_{OIII}$, we adopted the average value in the
following three bins, $0\leq R \leq 20$ arcsec, $20 < R \leq 40$
arcsec, $40< R \leq 60$ arcsec. The distribution of $t_3$ in each bin
is shown in Fig. \ref{t2t3}: the electron temperature is almost
constant till about 40 arcsec and tends to increase at larger
distances. In the same figure there are also plotted the values of
$t_2$ (at the same distances) obtained by adopting the following
empirical $t_2 - t_3$ relation, derived by \citet{Pil09}:
$t_2=0.264+0.835t_3$. The empirical $t_2 - t_3$ relation has been
investigated and debated for a long time and several forms are given
in the literature; that by \citet{Pil09} is the most recent work on
this subject, where they derived the $t_2 - t_3$ relation in terms of
the nebular $R_3 = [OIII]\lambda(4959+5007)/H_{\beta}$ and $R_2 =
[OII]\lambda3727/H_{\beta}$ line fluxes. They found that such relation is
valid for HII regions with a weak nebular $R_3$ lines ($logR_3 \ge
0.5$) and it turns to be consistent with those commonly used (see
Fig. 14 in \citet{Pil09} paper and reference therein). The polar disk
of NGC4650A, since $logR_3 = 0.5 \pm 0.2$ (this is the average value
on the whole disk extension), is located in the upper validity limit
of the $t_2 - t_3$ relation.\\ This is what is usually done since the
auroral line $[OII]\lambda7325$ is too weak to obtain a good enough
estimate of the electron temperature $t_2$. For the polar disk of
NGC4650A, the line fluxes of $[OII]\lambda(7320+7330)$ are strong enough
to be measured at about 30 arcsec and the average value of
$[OII]\lambda(7320+7330)/H_{\beta} = 0.04 \pm 0.01$. The average value
of the ratio $R_{OII} = [OII]\lambda3727/[OII]\lambda7325$ is given as input in the
TENDEM task, and we obtain $t_2 = (1.092 \pm 0.20) \times 10^{4}$ K,
which is added to Fig. \ref{t2t3}. This value is lower than that
derived by using photoionization models, even if the observed and
theoretical values of $t_2$ are consistent within errors: such
difference could be due both to the uncertainties in the line fluxes
measurements, made on so few and weak lines, and to the validity limit
of the $t_2 - t_3$ relation.\\ To obtain the oxygen abundance for the
polar disk, we decided to adopt both estimates for $t_2$, by using the
average value inside 40 arcsec for that derived by photoionization
models, which is $t_2^{mod} = (1.7 \pm 0.2) \times 10{^4}$ K, and,
consistently, an average value for $t_3 = (2.0 \pm 0.2) \times 10{^4}$
K at the same distances.  By adopting the Eq. \ref{OHIII} and
Eq. \ref{OHII}, the oxygen abundance $12+log(O/H)_{T}$ for the polar
disk is shown in Fig. \ref{OH_T} and listed in Tab. \ref{OHT}. By
using the value of $t_2$ derived by direct measurements of OII line
fluxes, the metallicity is higher with respect to that derived by
using the value of $t_2$ from photoionization models: the average
value is $12+log(O/H)_{T_{obs}} = 8.4 \pm 0.1$ and
$12+log(O/H)_{T_{mod}} = 7.6 \pm 0.5$ for the two cases respectively.
The large error, that are derived by propagating the emission line
intensity errors listed in Tab. \ref{fluxopt} and Tab. \ref{fluxnir},
which is about 0.5, let both estimates consistent.\\ It is important
to point out that the oxygen abundance derived by direct measurements
of OII line fluxes for $t_2$ is much more similar to the value derived
by empirical methods, which is $12+log(O/H) = 8.3$, than the value
derived by using photoionization models for $t_2$: this may suggest
that NGC4650A could be out of the range of validity of these models.

%% This section contains more display math examples, including unnumbered
%% equations (displaymath environment). The last paragraph includes some
%% examples of in-line math featuring a couple of the AASTeX symbol macros.

%% The displaymath environment will produce the same sort of equation as
%% the equation environment, except that the equation will not be numbered
%% by LaTeX.

%__________________________________________________________________

\section{Discussion: use of the chemical analysis to constrain the galaxy
formation}\label{disc}

%In order to test the cold accretion scenario for the formation of
%polar disk galaxy NGC4650A, we have used deep longslit spectroscopy,
%obtained with FORS2@VLT in the optical (078.B-0580) and NIR
%wavelengths (079.B-0177), to study the brightest HII regions
%associated with the polar disk and to derive the chemical abundances
%and SFR.
As discussed above, we derive the oxygen abundance, $12+log(O/H)$ to be $8.2 \pm 0.1$, by using
both the empirical and direct methods. In the
following sections we will discuss the results obtained by the present
work, how they could reconcile with the predictions by theoretical
models and, finally, we address the main conclusions of this study.

\subsection{Results}\label{res}
Here we discuss how the average value of metallicity and its
distribution along the polar disk compares with those typical for other
late-type disk galaxies and PRGs.

{\it - Metallicity-Luminosity relation}\\ The mean values for the
oxygen abundance along the polar disk, derived by the empirical (see
Sec. \ref{Emp}) and direct methods (see Sec. \ref{dir}), are compared with
those for a sample of late-type disk galaxies by \citet{Kob99}, as
function of the total luminosity (see Fig. \ref{conf}). For all
galaxies of the sample, and in particular for spiral galaxies, the
total luminosity is that of the whole system, i.e. bulge plus disk,
therefore also for NGC4650A we have accounted for the total luminosity
of the galaxy ($M_B=-19.3$, evaluated by using the same value of $H_{0}$
used by \citet{Kob99} in order to compare NGC4650A with galaxies in their
sample), to which contributes both the central
spheroid and the polar disk. We found that NGC4650A is located inside
the spread of the data-points and contrary to its high luminosity it
has metallicity lower than spiral galaxy disks of the same total
luminosity. If we take into account the total luminosity of the polar
disk alone ($M_B=-17$), NGC4650A falls in the region where HII and
irregular galaxies are also found, characterized by lower luminosity
and metallicities with respect to the spiral galaxies.\\ For what
concerns the chemical abundances in PRGs, only few measurements are
available in the literature. By using the direct method, very recently
\citet{Per09} have derived the chemical abundances of IIZw71, a blue
compact dwarf galaxy also catalogued as a probable polar ring:
consistently with its low luminosity, the metallicity of the brightest
knots in the ring is lower with respect to that of NGC4650A. Through a
similar study, \citet{Bro09} have derived the chemical
abundances for the apparent ring galaxy SDSS J075234.33+292049.8,
which has a more similar morphology to that observed in NGC4650A: the
average value for the oxygen abundance along the polar structure is
$12+logO/H = 8.49 \pm 0.08$. As pointed out by authors, taking into
account the ring brightness, such value is somewhat lower than that
expected by the metallicity-luminosity relation. A spectroscopic study
of the peculiar galaxy UGC5600, classified as probable PRG \citep{Whi90},
by \citet{Sha02} reports a quasi-solar
metallicity for such object, $12+log(O/H) \sim 8.8 = 0.9Z_{\odot}$, which
is consistent with its high luminosity ($M_B=-19.4$), but it is larger
with respect to the values of other PRGs given before.

\begin{figure*}
\centering
\includegraphics[width=11cm]{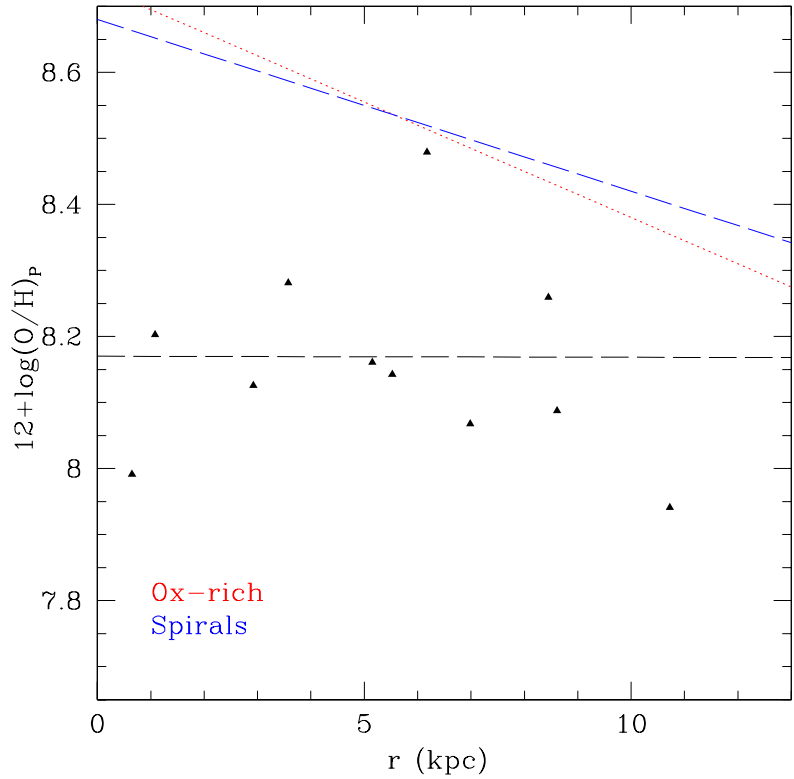}
\caption{Oxygen abundance derived with empirical methods proposed by
  \citet{Pil01} versus radius. The values in Tab. \ref{rs23} were
  binned each 5 arcseconds. The superimposed lines are the linear best
  fit derived by \citet{Pil06}; the red line represents the best fit
  to the abundance of oxygen-rich spirals, while the blue line is
  those related to ordinary spirals. The black line is the best fits
  obtained for NGC4650A.} \label{fit}
\end{figure*}

\begin{figure*}
\centering
\includegraphics[width=12cm]{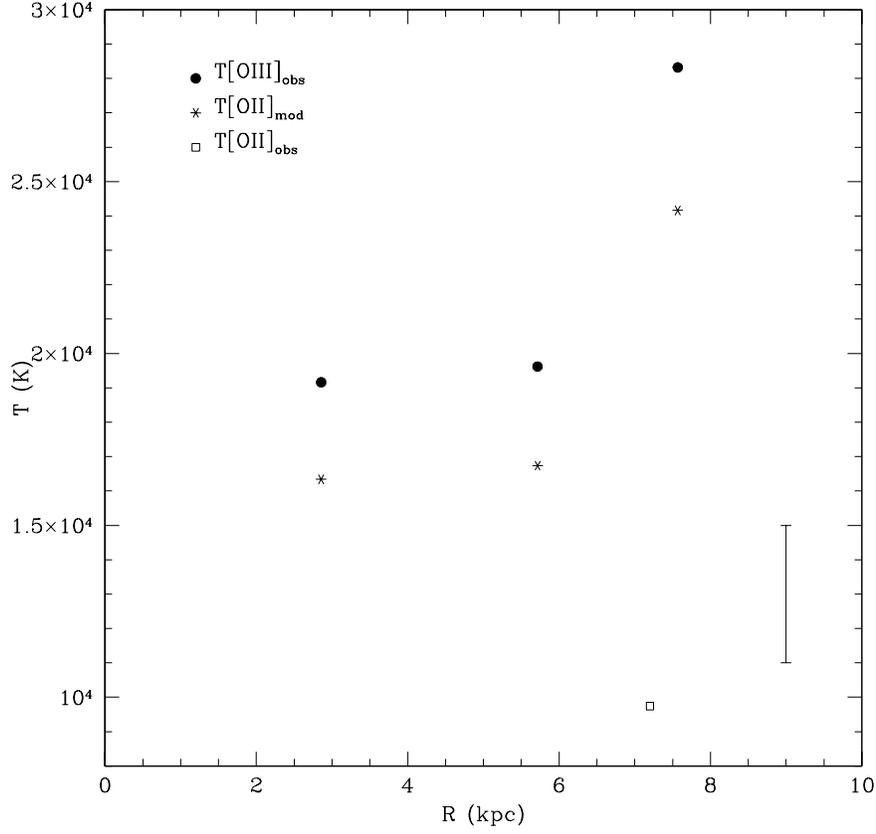}
\caption{Electron temperature along the polar disk of NGC4650A. The
  filled circles correspond to the observed T[OIII], the starred
  points correspond to the T[OII] estimated by the photoionization
  model (see Sec. \ref{dir} for details), the open square corresponds
  to the average value of T[OII] directly estimated by the ratio
  $R_{OII} = OII[3727]/[7325]$. In the bottom left side of the plot
  the mean error is shown.} \label{t2t3}
\end{figure*}

\begin{figure*}
\centering
\includegraphics[width=12cm]{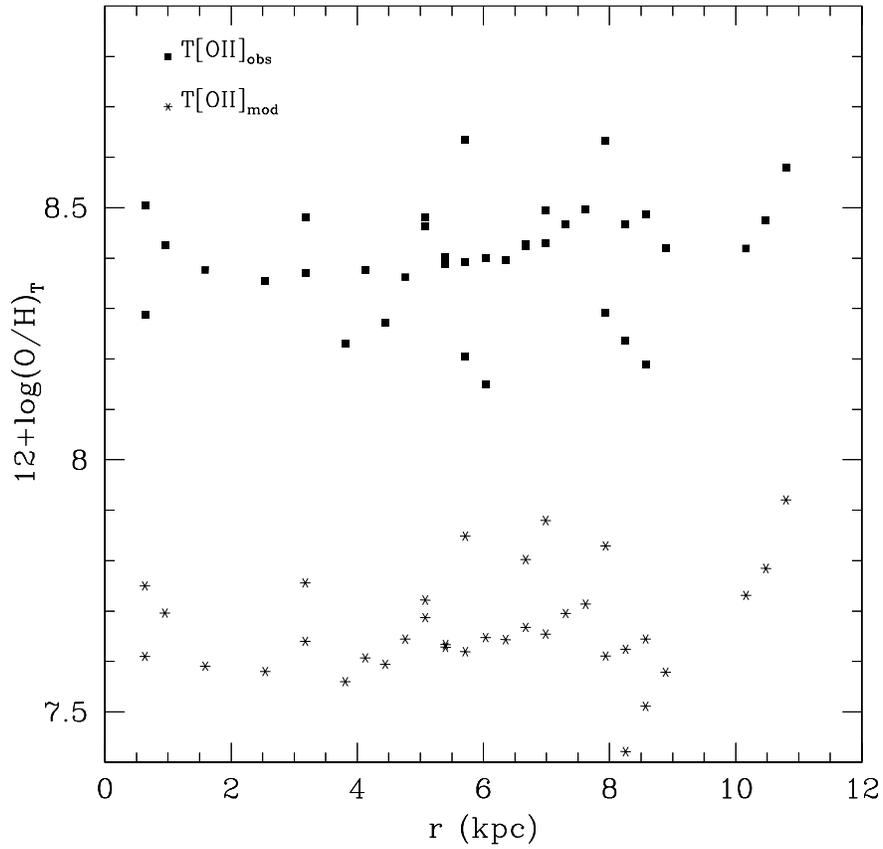}
\caption{Oxygen abundance derived with direct method. Squares are the
  values derived by using the observed $T[OII]$ estimates; stars are
  those derived by using the $T[OII]$ from photoionization models. See
  Sec. \ref{dir} for details. The errors on oxygen abundance is
  0.5.} \label{OH_T}
\end{figure*}

{\it - Metallicity gradient along the polar disk}\\ The oxygen
abundance in the polar disk of NGC4650A as a function of the radius
derived by empirical and direct methods are shown in Fig. \ref{fit}
and Fig. \ref{OH_T} respectively: both of them show that metallicity
remains constant along the polar disk. This suggests that the star
formation and metal enrichment in the polar structure of NGC4650A is
not influenced by the stellar evolution of the older central spheroid,
where the last burst of star formation occurred between 3 to 5 Gyrs
ago \citep{Iod02}: this turns also to be consistent with a later
formation of the polar disk. The absence of any metallicity gradient
is a typical behavior found in LSB galaxies \citep{deB98}, and also in
the polar disk galaxy studied by \citet{Bro09}, which has a very
similar structure to NGC4650A. On the contrary, ordinary and
oxygen-rich spiral galaxies show a decreasing abundance with
increasing radii (see Fig. \ref{fit} and \citealt{Pil06}). These
observed features in spiral disks are well explained by the infall
models of galaxy formation which predict that they build up via
accretion by growing inside-out (\citealt{Mat89}; \citealt{Boi99}) and
such process generates the observed gradients.

{\it - Star Formation Rate (SFR)}\\ We have derived the SFR for the
polar disk, from the $H_{\alpha}$ luminosity using the expression
given by \citet{Ken98} $SFR (M_{\odot}/yr)= 7.9 \times 10^{-42} \times
L(H_{\alpha}) (erg/s)$. We found that it is almost constant along the disk,
within a large scatter of the datapoint. Therefore, given the average
value of $L(H_{\alpha}) \simeq 3.8 \times 10^{39}$ erg/s we have
obtained an average $SFR \sim 0.04 M_{\odot}/yr$ and $SFR \sim 0.02
M_{\odot}/yr$ for the North and South arms respectively. These values
are quite similar to the total SFR derived for the HII regions of PRG
IIZw71 (\citealt{Per09}), which is $SFR \sim 0.035 M_{\odot}/yr$.  In
order to make a comparison with SFR estimates for a sample of PRGs
studied by \citet{Res94}, we have derived a mean value of the
$H_{\alpha}$ flux by integrating within a rectangular aperture of
$2.5'' \times 4.6''$ along each arm of the polar disk and we have
obtained an average value of $SFR \sim 0.006
M_{\odot}/yr/pc^{2}$. This value is consistent with the SFR of the HII
regions in the PRGs studied by \citet{Res94}, where $0.001 \le SFR \le
0.31$. In particular, two PRGs of this sample, UGC7576 and UGC9796,
where the polar structure is very similar to that of NGC4650A
(i.e. exponential surface brightness profile, very blue colors, knotty
appearance, prominent HII regions and a large amount of HI gas, all
associated to this component and distributed as a disk in differential
rotation; see \citealt{Res94} and \citealt{Cox06}), have $SFR \sim
0.005 M_{\odot}/yr/pc^{2}$. As already stressed by \citet{Res94}, the
HII regions of the PRGs, including also NGC4650A, have SFR similar to
late-type spiral galaxies \citep{Ken87}. In particular, by comparing
the $H_{\alpha}$ luminosity versus the total B magnitude, the polar
disk galaxies, NGC4650A, UGC7576 and UGC9796, are located in the same
area where Sc-Irr galaxies are also found.

Taking into account that the polar disk is very young, since the
  last burst of star formation occurred less than 1 Gyr ago
  \citep{Iod02}, we check if the present SFR of 0.06 $M_{\odot}/yr$,
  and even 2 and 3 times higher (i.e. $SFR = 0.12 M_{\odot}/yr$ and
  $SFR = 0.18 M_{\odot}/yr$), can give the estimated metallicity of
  $Z=0.2Z_{\odot}$ and how strongly could increase the metallicity
  with time.  We used a linearly declining SFR \citep{Bru03} $\psi(t)
  = 2M_{\star}\tau^{-1}[1-(t/\tau)]$ (typically used for late-type
  galaxies), to estimate the expected stellar mass for the three
  different values of the SFR (0.06, 0.12 and 0.18 $M_{\odot}/yr$) and
  three epochs (0.8 Gyr, 1 Gyr and 2 Gyrs), obtaining stellar masses
  in the range $4 \times 10^{9} M_{\odot} \leq M_{\star} \leq 1 \times
  10^{10} M_{\odot}$. The stellar
  mass ($M_{\star} \sim 4 \times 10^{9} M_{\odot}$) in the disk from
  NIR observations \citep{Iod02} falls within this range. Than, by using the
  mass-metallicity relation derived by \citet{Tre04}, where
  $12+log(O/H) = -1.492 + 1.847 log(M_{\star}) - 0.08026 (log
  M_{\star})^{2}$, we found that $1.02 Z_{\odot} \le Z \le 1.4
  Z_{\odot}$. This shows that the present SFR for the polar disk ($SFR
  = 0.06 M_{\odot} yr^{-1}$) is able to increase the metallicity of
  about $0.2 Z_{\odot}$ after 1Gyr. The derived values for Z are
  larger than $Z=(0.2 \pm 0.002) Z_{\odot}$, found by using the
  element abundances (see Sec.\ref{Emp}): this differences could be
  attributed to the accretion of metal-poor gas, as discussed in
  detail in the following sections.

\subsection{Theoretical predictions}\label{theory}

How galaxies acquire their gas is an open issue in the models of
galaxy formation and recent theoretical works have argued that cold
accretion plays a major role. This idea is supported by many numerical
simulations suggesting that this could be an important mechanism in
the galaxy formation (\citealt{Kat93}; \citealt{Kat94};
\citealt{Ker05}; \citealt{Dek06}; \citealt{Dek08}; \citealt{Bou09}).
\citet{Ker05} studied in detail the physics of the {\it cold mode}
of gas accretion and one feature they found is that it is generally
directed along filaments, allowing galaxies to draw gas from large
distances.\\ Recently \citet{Ocv08}, using cosmological simulations,
studied the metal enrichment of the intergalactic medium as a key
ingredient in determining the transition mass from cold to hot
dominated accretion.  Their measurements turn to be consistent with
the analytical prediction of \citet{Dek06} in the low metallicity
regime for the cold streams. The efficiency of radiative cooling
decreases towards low metallicity, and at $Z/Z_{\odot} = 10^{-3}$ the
cooling properties of the gas are those of a primordial mixture. Gas
accretion is bimodal both in temperature and in metallicity: the cold
accretion mode is associated with a combination of metal-poor
filamentary accretion and dense metal-rich satellite galaxy disc
surrounding, while the hot mode features has strong chemical
heterogeneity and a radius-dependent metallicity.

More recent simulations of disk formation in a cosmological
context performed by \citet{Age09} revealed that the so called
chain-galaxies and clump-clusters, found only at higher
redshifts \citep{Elm07}, are a natural outcome of early epoch
enhanced gas accretion from cold dense streams as well as tidally
and ram-pressured stripped material from minor mergers and
satellites.  This freshly accreted cold gas settles into a large
disk-like systems; simulations show that the interaction region
between the new-formed disk and the cold streams can also result
not aligned with the initial galactic disk: based on a very
poor statistics, \citet{Age09} suggest that this misalignment
might not be typical, and it is due to a third cold stream that is
perpendicular to the main filament. A more recent analysis by
\citet{Dek09N} shows that the accretion of gas along misaligned
filaments with respect to the disk plane are more common and it
leaves traces down to low redshift. An almost polar filament can
result just as an extreme case of such process and, as suggested
by \citet{Age09}, it could be responsible for the formation of
polar disks.\\ This scenario not only reproduces the observed
morphology and global rotation of disks, but also finds a realistic
metallicity gradient and SFR of $20 M_{\odot}/yr$. \citet{Age09}
found solar metallicity for the inner disk, while that in the
clump forming region is only $\sim 1/10 Z_{\odot}$ due to the
accretion of pristine gas in the cold streams mixing with stripped
satellite gas. Therefore, the studies cited above suggest that the
chemical abundance is one of the key parameters that can be
estimated in a galaxy disk and directly compared with the
theoretical predictions. This is the main goal of the present
work: we studied the chemical abundances in the polar disk of
NGC4650A in order to check the cold accretion scenario for this
object.

Hydrodynamical simulations performed by \citet{Mac06} and
\citet{Bro08} have shown that the formation of a polar disk galaxy can
occur naturally in a hierarchical universe, where most low-mass
galaxies are assembled through the accretion of cold gas infalling
along a filamentary structures. According to \citet{Mac06}, the polar
disk forms from cold gas that flows along the extended $\sim 1 Mpc$
filament into the virialized dark matter halo. The gas streams into
the center of the halo on an orbit that is offset from radial
infall. As it reaches the center, it impacts with gas in the halo of
the host galaxy and with the warm gas flowing along the opposite
filament. Only the gas accreted perpendicular to the major axis of the
potential can survive for more than a few dynamical lifetimes.\\
\citet{Bro08} argued that polar disk galaxies are extreme examples of
the misalignment of angular momentum that occurs during the
hierarchical structure formation: an inner disk starts forming shortly
after the last major merger at $z \sim 2$. Due to its gas rich nature,
the galaxy rapidly forms a new disk whose angular momentum is
determined by the merger orbital parameters. Later, gas continues to
be accreted but in a plane that is almost perpendicular to the inner
disk. At $z \sim 0.8$ the central galaxy is still forming stars in a
disk, while the bulk of new star formation is in the highly inclined
polar disk. By $z\sim 0.5$ the inner disk has exhausted its gas, while
gas continues to fall onto the polar disk. From this point, star
formation occurs exclusively in the polar disk, that remains stable
for at least 3 Gyrs.\\ The formation mechanisms described above can
self-consistently explain both morphology and kinematics of a polar
disk galaxy and, in particular, all the observed features (like colors
and color gradients, longevity, spiral arms, HI content and
distribution) of the polar structure.

\section{Summary and conclusions}\label{concl}

The present study could be considered a step forward both to trace the
formation history of NGC4650A and to give hints on the mechanisms at
work during the building of a disk by cold accretion process. 

As mentioned in the Sec.\ref{intro}, the new kinematic data
  obtained for the central spheroid \citep{Iod06}, together with the
  previous studies, set important constraints on the possible
  formation mechanisms for NGC4650A.  
In particular, the merging scenario is ruled out because,
  according to simulations (e.g. \citealt{Bou05}), a high mass ratio
  of the two merging galaxies is required to form a massive and
  extended polar disk as observed in NGC~4650A: this would convert the
  intruder into an elliptical-like, not rotationally supported,
  stellar system. This is in contrast with the high maximum rotation
  velocity ($\sim 80 \div 100$ km/s) observed in the outer regions of
  the central spheroid \citep{Iod06}.

 Both the high baryonic mass (star plus gas) in the polar structure
 and its large extension cannot reconcile with a polar ring formed via
 the gradual disruption of a dwarf satellite galaxy (as explained by
 \citealt{Iod02}).  Differently, a wide polar ring and/or disk (as
 observed in NGC4650A) may form both around a disk or an elliptical
 galaxy through the tidal accretion of gas stripped from a gas-rich
 donor, in the case of large relative velocity and large impact
 parameter and for a particular orbital configuration
 (e.g. \citealt{Bou03}).  Therefore, the two formation scenarios which
 can be really envisioned in the specific case of NGC4650A are the
 tidal accretion and the accretion of external primordial cold gas
 from cosmic web filaments (\citealt{Mac06}; \citealt{Bro08}). To this
 aim we have derived the metallicity and SFR for the polar disk in
 NGC4650A in order to compare them with those predicted by different
 formation scenarios.

\begin{figure*}
\centering
\includegraphics[width=12cm]{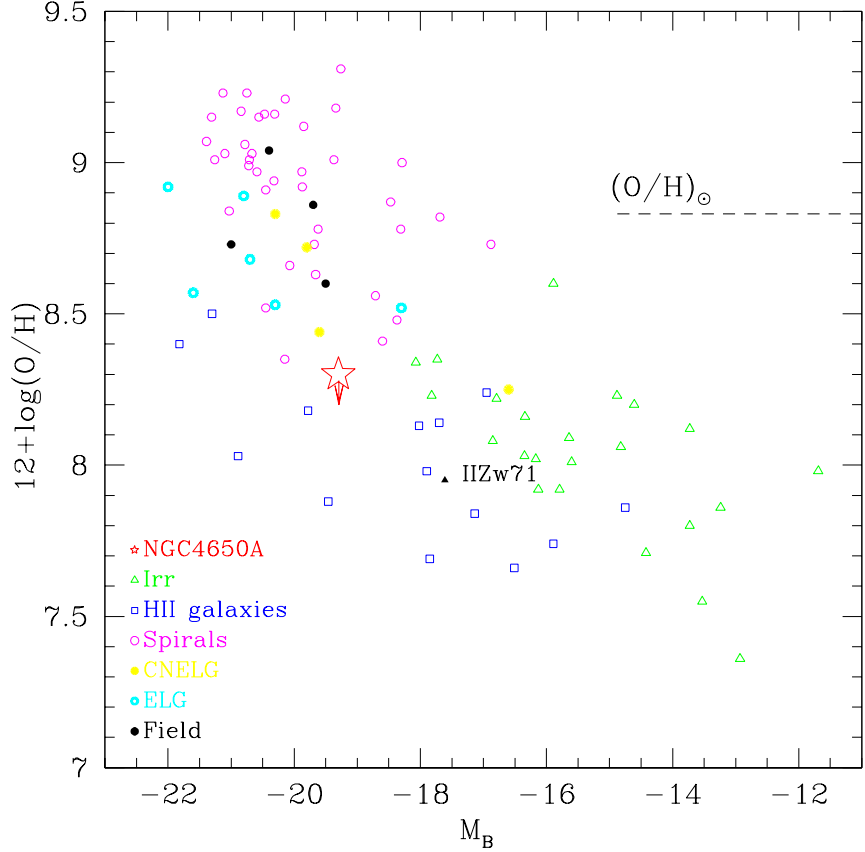}
\caption{Oxygen abundance vs absolute blue magnitude for CNELGs
  (yellow filled circles), ELGs (cyan open circles), four field
  galaxies with emission lines (filled black circles), nearby dwarf
  irregulars (open triangles), local spiral galaxies (open circles),
  local HII galaxies (open squares), NGC4650A (star) and the polar
  disk galaxy IIZw71 \citep{Per09}. The dashed line indicates the
  solar oxygen abundance. The arrow indicates the shift of the value
  of the oxygen abundance if we use the direct methods to evaluate
  it. The total B-band magnitude for NGC4650A ($M_B=-19.3$) has been
  evaluated by using the same value of $H_{0}$ used by \citet{Kob99}
  in order to compare NGC4650A with galaxies in their
  sample} \label{conf}
\end{figure*}

The main results of the
  present work are (see also Sec. \ref{res}): \emph{i}) the low value
  of the metallicity derived for the polar disk $Z = 0.2Z_{\odot}$, in
  spite of its high luminosity, $M_{B} = -19.3$ (see Fig. \ref{conf}),
  \emph{ii}) the lack of any metallicity gradient along the polar
  disk, which suggests that the metal enrichment is not influenced by
  the stellar evolution of the older central spheroid, \emph{iii}) the
  metallicities expected for the present SFR at three different
  epochs, $1.02 Z_{\odot} \le Z \le 1.4 Z_{\odot}$, are higher than
  those measured from the element abundances and this is consistent
  with a later infall of metal-poor gas.

In the following we will address how these results reconcile with the
predictions by theoretical models (see Sec. \ref{theory}) and may
discriminate between the two formation mechanisms.

If the polar ring/disk formed by the mass transfer from a gas-rich
donor galaxy, the accreted material comes only from the outer and more
metal-poor regions of the donor: is the observed metallicity for the
polar component in NGC4650A consistent with the typical values for the
outer disk of a bright spiral galaxy? According to \citet{Bre09}, the
metallicity of very outer regions of a bright spiral is $0.2 Z_{\odot} \le Z \le
1.1 Z_{\odot}$: the observed value for NGC4650A, $Z=0.2Z_{\odot}$, is
close to the lower limit.\\ The cold accretion mechanism for disk
formation predicts quite low metallicity ($Z=0.1Z_{\odot}$)
(\citealt{Dek06}, \citealt{Ocv08}, \citealt{Age09}): such value refers
to the time just after the accretion of a misaligned material, so it
can be considered as initial value for Z before the subsequent
enrichment. How this may reconcile with the observed metallicity for
NGC4650A?  We estimated that the present SFR for the polar disk ($SFR
= 0.06 M_{\odot} yr^{-1}$) is able to increase the metallicity of
about 0.2 after 1Gyr (see Sec.\ref{res}): taking into account that the
polar structure is very young, less than 1Gyr \citep{Iod02}, an
initial value of $Z=0.1Z_{\odot}$, at the time of polar disk
formation, could be consistent with the today's observed
metallicity.\\ This evidence may put some constraints also on the
time-scales of the accretion process. The issue that need to be
addressed is: how could a cosmic flow form a ring/disk only in the
last Gyr, without forming it before?  Given that the average age of
0.8 Gyr refers to the last burst of star formation, reasonably the gas
accretion along the polar direction could have started much earlier
and stars formed only recently, once enough gas mass has been
accumulated in the polar disk. Earlier-on, two possible mechanisms may
be proposed. One process that may happen is something similar to that
suggested by \citet{Mar09} for the quenching of star formation in
early-type galaxies: given that star formation takes place in
gravitationally unstable gas disks, the polar structure could have
been stable for a while and this could have quenched its star
formation activity, until the accumulated mass gas exceeds a stability
threshold and star formation resumes. Alternatively, according to the
simulations by \citet{Bro08}, the filament could have been there for
several Gyrs with a relative low inclination, providing gas fuel to
the star formation in the host galaxy first, about 3 Gyrs ago. Then
large-scale tidal fields can let the disk/filament misalignment
increase over the time till 80-90 degrees and start forming the polar
structure during the last 1-2 Gyrs. This picture might be consistent
not only with the relative young age of the polar disk, but also with
the estimate of the last burst of star formation in the central
spheroid \citep{Iod02}.

One more hint for the cold accretion scenario comes from the fact
that the metallicity expected by the present SFR turns to be
higher than those directly measured by the chemical abundances. As
suggested by \citet{Dal07}, both infall and outflow of gas can
change a galaxy's metallicity: in the case of NGC4650A a possible
explanations for this difference could be the infall of pristine
gas, as suggested by (\citealt{Fin07}; \citealt{Ell08}).

The lack of abundance gradient in the polar disk, as typically
observed also in LSB galaxies, suggests that the picture of a chemical
evolution from inside-out, that well reproduce the observed features
in spiral disks (\citealt{Mat89}, \citealt{Boi99}), cannot be applied
to these systems.  In particular, observations for LSB galaxies are
consistent with a quiescent and sporadic chemical evolution, but
several explanation exist that supports such evidence
\citep{deB98}. Among them, one suggestion is that the disk is still
settling in its final configuration and the star formation is
triggered by external infall of gas from larger radii: during this
process, gas is slowly diffusing inward, causing star formation where
conditions are favorable. As a consequence, the star formation is not
self-propagating and the building-up of the disk would not give rise
to an abundance gradient. The similarities between LSB galaxies and
the polar disk in NGC4650A, including colors \citep{Iod02} and
chemical abundances (this work), together with the very young age of
the polar disk, the presence of star forming regions towards larger
radii, the warping structure of outer arms (\citealt{Gal02},
\citealt{Iod02}), and the constant SFR along the disk (this work),
suggest that the infall of metal-poor gas, through a similar process
described above, may reasonably fit all these observational evidences.

Given that, are there any other observational
aspects that can help to disentangle in a non ambiguous way the
two scenarios?

One important feature which characterize NGC4650A is the high
baryonic (gas plus stars) mass in the polar structure: it is about
$12 \times 10^{9} M_\odot$, which is comparable with or even
higher than the total luminous mass in the central spheroid (of
about $5 \times 10^{9} M_\odot$). In the accretion scenario, the
total amount of accreted gas by the early-type object is about
$10\%$ of the gas in the disk donor galaxy, i.e. up to $10^{9}$
$M_{\odot}$, so one would expect the accreted baryonic mass
(stellar + gas) to be a fraction of that in the pre-existing
galaxy, and not viceversa, as it is observed for NGC4650A.
Furthermore, looking at the field around NGC4650A, the close
luminous spiral NGC4650 may be considered as possible donor galaxy
\citep{Bou03}, however the available observations on the HI
content for this object show that NGC4650 is expected to be
gas-poor (\citealt{Arn97}; \citealt{Van02}). So, where the high
quantity of HI gas in NGC4650A may come from? If the polar disk
forms by the cold accretion of gas from filaments there is no
limit to the accreted mass.

Given all the evidences shown above, we can infer that the cold
  accretion of gas by cosmic web filaments could well account for both
  the low metallicity, the lack of gradient and the high HI content in
  NGC4650A. An independent evidence which seems to support such
  scenario for the formation of polar disks comes from the discovery
  and study of an isolated polar disk galaxy, located in a wall
  between two voids \citep{Sta09}: the large HI mass (at least
  comparable to the stellar mass of the central galaxy) and the
  general underdensity of the environment can be consistent with the
  cold flow accretion of gas as possible formation mechanism for this
  object.

The present work remarks how the use of the chemical analysis can give
strong constraints on the galaxy formation, in particular, it has
revealed an independent check of the cold accretion scenario for the
formation of polar disk galaxies. This study also confirmed that this
class of object needs to be treated differently from the polar ring
galaxies, where the polar structure is more metal rich (like UGC5600,
see Sec. \ref{res}) and a tidal accretion or a major merging process
can reliable explain the observed properties \citep{Bou03}.
Finally, given the similarities between polar disks and
late-type disk galaxies, except for the different plane with respect
to the central spheroid, the two classes of systems could share similar
formation processes. Therefore, the study of polar disk galaxies assumes
an important role in the wide framework of disk formation and
evolution, in particular, for what concern the ``rebuilding'' of disks
through accretion of gas from cosmic filaments, as predicted by
hierarchical models of galaxy formation \citep{Ste02}.


%% Included in this acknowledgments section are examples of the
%% AASTeX hypertext markup commands. Use \url without the optional [HREF]
%% argument when you want to print the url directly in the text. Otherwise,
%% use either \url or \anchor, with the HREF as the first argument and the
%% text to be printed in the second.

\acknowledgments

The authors wish to thank the referee, Frederic Bournaud, for the
detailed and constructive report, which allowed to improve the
paper. This work is based on observations made with ESO Telescopes at
the Paranal Observatories under programme ID $<078.B-0580(A)>$ and
$<079.B-0177(A)>$. M.S. and E.I. wish to thank the Max-Plank-Institut
f\"{u}r Extraterrestrische Physik for the hospitality given during
their work. M.S. wish to thank the University of Naples ''Federico
II'', the Astronomical Observatory of Capodimonte and the
Max-Plank-Institut f\"{u}r Extraterrestrische Physik for the financial
support given during this work. E.I wish to thank ESO for the
financial support and hospitality given during her visiting on June
2008. M.S. and E.I. are very grateful to C. Tortora for many useful
discussion and suggestions.

\clearpage

\begin{table}
\begin{minipage}[t]{\columnwidth}
\caption{General properties of NGC4650A}% title of Table
\label{General} % is used to refer this table in the text
\centering
\renewcommand{\footnoterule}{}  % used for centering table
\begin{tabular}{l l l}        % centered columns (4 columns)
\hline\hline                 % inserts double horizontal lines
Parameter & Value & Ref. \\    % table heading
\hline                        % inserts single horizontal line
   Morphological type & PRG &  NED\footnote{NASA/IPAC Extragalactic Database}\\       % inserting body of the table
   R.A. (J2000) & 12h44m49.0s & NED \\
   Dec. (J2000) & -40d42m52s & NED \\
   Helio. Radial Velocity & 2880 km/s & NED  \\
   Redshift & 0.009607 & NED     \\
   Distance & 38 Mpc &     \\
   Total m$_{B}$ (mag) & 14.09 $\pm$ 0.21 & NED \\
   $M(HI)(M_{\bigodot})$ & $8 \times\ 10^{9}$ & \cite{Arn97}\\
   $L_{B} (L_{\bigodot})$ & $3 \times\ 10^{9}$ & \\

\hline                      %inserts single line
\end{tabular}
\end{minipage}
\end{table}

\clearpage

\begin{deluxetable}{crrrrrrrrrrr}
\tablecolumns{11}
\tablewidth{0pc}
\tablecaption{Observed and de-reddened emission lines fluxes relative to
$H_{\beta}$ in the $3300-6210$ \AA\ wavelength range. \label{fluxopt}}
\tablehead{
\multicolumn{1}{c}{r (arcsec)\tablenotemark{a}}    &  \multicolumn{2}{c}{OII[3727]/$H_{\beta}$} &  \multicolumn{2}{c}{$H_{\gamma}$[4340]/$H_{\beta}$} & \multicolumn{2}{c}{OIII[4363]/$H_{\beta}$} & \multicolumn{2}{c}{OIII[4959]/$H_{\beta}$} & \multicolumn{2}{c}{OIII[5007]/$H_{\beta}$}}    % table heading
%\multicolumn{1}{c}{slit 1}& \multicolumn{1}{c}{$F_{obs}$} & \multicolumn{1}{c}{$F_{int}$} & \multicolumn{1}{c}{$F_{obs}$} & \multicolumn{1}{c}{$F_{int}$} & \multicolumn{1}{c}{$F_{obs}$} & \multicolumn{1}{c}{$F_{int}$} & \multicolumn{1}{c}{$F_{obs}$} & \multicolumn{1}{c}{$F_{int}$} & \multicolumn{1}{c}{$F_{obs}$} & \multicolumn{1}{c}{$F_{int}}$}
\startdata
slit1  & $F_{obs}$ & $F_{int}$ & $F_{obs}$ & $F_{int}$ & $F_{obs}$ & $F_{int}$ & $F_{obs}$ & $F_{int}$ &$F_{obs}$ & $F_{int}$\\
 & $\pm 0.05$ & $\pm 0.06$ & $\pm 0.013$ & $\pm 0.015$ & $\pm 0.004$ & $\pm 0.005$ & $\pm 0.02$ & $\pm 0.02$ & $\pm 0.03$ & $\pm 0.03$\\
\tableline
-44.1  & 5.95 & 7.76 & \nodata  & \nodata  & 0.058 & 0.066 & 0.52 & 0.51 & 1.83 & 1.77\\
-42.34 & 4.32 & 5.64 & \nodata  & \nodata  & 0.201 & 0.227 & 0.52 & 0.51 & 1.83 & 1.77\\
-40.57 & 4.01 & 5.22 & 0.392& 0.445& 0.182 & 0.205 & 0.58 & 0.57 & 1.97 & 1.91\\
-38.81 & 3.69 & 4.81 & \nodata  & \nodata  & 0.094 & 0.107 & 0.57 & 0.56 & 1.67 & 1.61\\
-37.04 & 3.61 & 4.71 & 0.438& 0.497& 0.096 & 0.109 & 0.80 & 0.78 & 1.91 & 1.85\\
-35.28 & 3.39 & 4.42 & 0.398& 0.451& 0.095 & 0.108 & 0.80 & 0.78 & 1.82 & 1.76\\
-33.52 & 3.42 & 4.45 & 0.329& 0.373& 0.064 & 0.073 & 0.79 & 0.78 & 1.87 & 1.81\\
-31.75 & 3.38 & 4.41 & 0.377& 0.428& 0.043 & 0.048 & 0.51 & 0.50 & 1.60 & 1.55\\
-29.99 & 3.47 & 4.52 & 0.423& 0.480& 0.038 & 0.043 & 0.54 & 0.52 & 1.58 & 1.53\\
-28.22 & 3.99 & 5.21 & 0.464& 0.526& 0.188 & 0.212 & 0.60 & 0.59 & 1.76 & 1.71\\
-17.64 & 4.06 & 5.29 & 0.400& 0.454& 0.105 & 0.119 & 1.05 & 1.03 & 3.11 & 3.01\\
-14.11 & 3.12 & 4.06 & 0.376& 0.427& \nodata   & \nodata   & 0.50 & 0.50 & 1.36 & 1.31\\
-3.528 & 4.35 & 5.67 & \nodata  & \nodata  & \nodata   & \nodata   & 0.81 & 0.80 & 2.51 & 2.43\\
26.46 & \nodata  & \nodata  & 0.348& 0.395& \nodata   & \nodata   & \nodata  & \nodata  & \nodata  & \nodata\\

28.22 & \nodata  & \nodata  & 0.540& 0.613& \nodata   & \nodata   & \nodata  & \nodata  & \nodata  & \nodata\\
37.04 & 3.39 & 4.42 & \nodata  & \nodata  & \nodata   & \nodata   & 1.79 & 1.75 & 5.33 & 5.16\\
38.81 & 3.94 & 5.13 & \nodata  & \nodata  & \nodata   & \nodata   & 1.92 & 1.88 & 6.85 & 6.63\\
45.86 & 2.41 & 3.14 & \nodata  & \nodata  & \nodata   & \nodata   & 0.24 & 0.24 & 0.51 & 0.49\\
47.63 & 4.33 & 5.64 & \nodata  & \nodata  & 0.030 & 0.034 & 0.21 & 0.20 & 0.37 & 0.36\\
49.39 & 3.70 & 4.82 & \nodata  & \nodata  & \nodata   & \nodata   & 0.20 & 0.20 & 0.33 & 0.32\\
 & & & & & & & & & & \\
slit 2 & & & & & & & & & & \\
 & & & & & & & & & & \\
-49.49 & \nodata   & \nodata & 0.598& 0.678& \nodata  & \nodata  & \nodata  & \nodata  & \nodata  & \nodata\\
-33.52 & 1.97 & 2.57 & 0.555& 0.630& \nodata  & \nodata  & 0.19 & 0.19 & 0.43 & 0.42\\
-31.75 & 2.24 & 2.91 & \nodata  & \nodata  & \nodata  & \nodata  & 0.17 & 0.17 & 0.54 & 0.53\\
-3.528 & 2.54 & 3.31 & \nodata  & \nodata  & \nodata  & \nodata  & 0.88 & 0.86 & 2.89 & 2.80\\
5.292 & 3.59 & 4.69 & 0.233& 0.264& \nodata  & \nodata  & 0.80 & 0.78 & 2.69 & 2.60\\
8.82  & 3.28 & 4.28 & \nodata  & \nodata  & \nodata  & \nodata  & 0.43 & 0.42 & 1.27 & 1.23\\
17.64 & 3.15 & 4.11 & 0.594& 0.674& \nodata  & \nodata  & 0.92 & 0.90 & 2.15 & 2.08\\
21.17 & 2.22 & 2.90 & \nodata  & \nodata  & \nodata  & \nodata  & 0.88 & 0.86 & 2.56 & 2.48\\
22.93 & 3.27 & 4.26 & 0.386& 0.438& \nodata  & \nodata  & 0.51 & 0.50 & 1.58 & 1.53\\
24.7  & 2.44 & 3.19 & 0.387& 0.439& \nodata  & \nodata  & 0.93 & 0.91 & 2.71 & 2.62\\
26.46 & 3.08 & 4.01 & 0.442& 0.501& \nodata  & \nodata  & 0.83 & 0.82 & 2.51 & 2.43\\
28.22 & 4.13 & 5.38 & 0.444& 0.504& \nodata  & \nodata  & 0.75 & 0.73 & 2.27 & 2.19\\
29.99 & 3.32 & 4.33 & 0.472& 0.535& \nodata  & \nodata  & 0.64 & 0.62 & 1.91 & 1.85\\
31.75 & 5.96 & 7.77 & 0.423& 0.480& \nodata  & \nodata  & 0.83 & 0.81 & 2.22 & 2.15\\
33.51 & \nodata  & \nodata  & 0.540& 0.613& \nodata  & \nodata  & \nodata  & \nodata  & \nodata  & \nodata\\
44.1  & 2.57 & 3.35 & \nodata  & \nodata  & \nodata  & \nodata  & 0.88 & 0.86 & 2.84 & 2.75\\
45.86 & 4.14 & 5.40 & 0.427& 0.484& 0.129& 0.146& 0.09 & 0.09 & 0.43 & 0.42\\
47.63 & 2.02 & 2.64 & \nodata  & \nodata  & 0.537& 0.606& 0.73 & 0.71 & 2.27 & 2.19\\
49.39 & \nodata  & \nodata  & 0.384& 0.436& \nodata  & \nodata  & \nodata  & \nodata  & \nodata  & \nodata\\
56.45 & 3.46 & 4.51 & \nodata  & \nodata  & \nodata  & \nodata  & 1.04 & 1.02 & 3.72 & 3.60\\
58.21 & 3.93 & 5.13 & \nodata  & \nodata  & 0.162& 0.183& 1.15 & 1.13 & 4.18 & 4.05\\
59.98 & 4.92 & 6.41 & \nodata  & \nodata  & \nodata  & \nodata  & 1.68 & 1.65 & 6.63 & 6.42\\
\enddata
\tablenotetext{a}{The negative values are for the northern regions of the spectra}
\end{deluxetable}

\clearpage

\begin{deluxetable}{crrrrrrrrrrr}
\tablecolumns{11}
\tablewidth{0pc}
\tablecaption{Observed and de-reddened emission lines fluxes relative to
$H_{\beta}$ in the $5600-11000$\AA\ wavelength range. \label{fluxnir}}
\tablehead{
\multicolumn{1}{c}{r (arcsec)\tablenotemark{a}}    &  \multicolumn{2}{c}{$H_{\alpha}[6563]/H_{\beta}$\tablenotemark{b}} &  \multicolumn{2}{c}{SII[6717]/$H_{\beta}$} & \multicolumn{2}{c}{SII[6731]/$H_{\beta}$}  & \multicolumn{2}{c}{SIII[9069]/$H_{\beta}$} & \multicolumn{2}{c}{SIII[9532]/$H_{\beta}$}}    % table heading
%\multicolumn{1}{c}{slit 1}& \multicolumn{1}{c}{$F_{obs}$} & \multicolumn{1}{c}{$F_{int}$} & \multicolumn{1}{c}{$F_{obs}$} & \multicolumn{1}{c}{$F_{int}$} & \multicolumn{1}{c}{$F_{obs}$} & \multicolumn{1}{c}{$F_{int}$} & \multicolumn{1}{c}{$F_{obs}$} & \multicolumn{1}{c}{$F_{int}$} & \multicolumn{1}{c}{$F_{obs}$} & \multicolumn{1}{c}{$F_{int}}$}
\startdata
 slit 1 & $F_{obs}$ & $F_{int}$ & $F_{obs}$ & $F_{int}$ & $F_{obs}$ & $F_{int}$ & $F_{obs}$ & $F_{int}$ & $F_{obs}$ & $F_{int}$ \\
  & $\pm 0.02$ & $\pm 0.03$ & $\pm 0.015$ & $\pm 0.011$ & $\pm 0.016$ & $\pm 0.012$ & $\pm 0.08$ & $\pm 0.06$ & $\pm 0.15$ & $\pm 0.11$\\
\tableline
-38.81 &\nodata &\nodata &\nodata &\nodata &\nodata &\nodata &\nodata &\nodata &0.05  &0.03  \\
-37.04 &5.46  &4.25  &\nodata &\nodata &\nodata &\nodata & \nodata &\nodata &\nodata &\nodata \\
-35.28 &3.85  &2.99  &\nodata &\nodata &\nodata &\nodata & \nodata &\nodata &0.017  &0.010  \\
-33.52 &2.44  &1.896  &\nodata &\nodata &\nodata &\nodata &\nodata &\nodata &0.04  &0.03 \\
-31.75 &0.930  &0.724 &\nodata &\nodata &\nodata &\nodata &\nodata &\nodata &\nodata &\nodata \\
-29.99 &0.412  &0.321  &\nodata &\nodata &\nodata &\nodata &\nodata &\nodata &\nodata &\nodata \\
-26.46 &\nodata &\nodata &\nodata &\nodata &\nodata &\nodata  &\nodata &\nodata &0.05 &0.03 \\
-17.64 &\nodata &\nodata &\nodata &\nodata &\nodata &\nodata  &\nodata &\nodata &0.03 &0.019 \\
-15.88 &\nodata &\nodata &\nodata &\nodata &\nodata &\nodata &\nodata &\nodata &0.06 &0.03 \\
-10.58 &\nodata &\nodata &0.199 &0.153 &0.632 &0.483 &\nodata &\nodata &\nodata &\nodata \\
-8.82  &\nodata &\nodata &0.426 &0.329 &0.577 &0.441 &0.052 &0.0320 &\nodata &\nodata\\
-7.06  &\nodata &\nodata &0.84 &0.646 &0.935 &0.714 &0.0293 &0.0178 &\nodata &\nodata\\
-5.29 &\nodata &\nodata &\nodata &\nodata  &\nodata &\nodata &0.041 &0.0252 &\nodata &\nodata \\
-3.53 &\nodata &\nodata &\nodata &\nodata  &\nodata &\nodata &0.68 &0.41 &\nodata &\nodata \\
-1.76 &\nodata &\nodata &\nodata &\nodata  &\nodata &\nodata &0.266 &0.162 &\nodata &\nodata \\
0.0  &\nodata &\nodata &1.39 &1.07 &1.31 &1.00 &\nodata &\nodata &\nodata &\nodata \\
1.76 &\nodata &\nodata &1.48 &1.14 &0.609 &0.465 &0.33 &0.202 &\nodata &\nodata \\
3.53 &\nodata &\nodata &0.686 &0.529 &0.460 &0.351 &\nodata &\nodata &\nodata &\nodata \\
5.29 &\nodata &\nodata &0.136 &0.105 &0.422 &0.322 &\nodata &\nodata &\nodata &\nodata \\
7.06 &\nodata&\nodata&0.0142 &0.0109 &0.377 &0.288 &0.041 &0.0249 &0.023 &0.014 \\
29.99 &\nodata &\nodata&\nodata &\nodata &\nodata &\nodata &\nodata &\nodata &0.014 &0.008 \\
31.75 &\nodata &\nodata&\nodata &\nodata &\nodata &\nodata &\nodata &\nodata &0.022 &0.013  \\
40.57 &\nodata &\nodata&0.0340 &0.0262 &0.174 &0.133 &\nodata &\nodata &\nodata &\nodata \\
51.16 &\nodata &\nodata&0.465 &0.358 &0.280 &0.214 &0.33 &0.202 &1.0 &0.6 \\
 & & & & & & & & & &  \\
slit 2 & & & & & & & & & &  \\
 & & & & & & & & & & \\
-45.86 &\nodata &\nodata &\nodata &\nodata &\nodata &\nodata &\nodata &\nodata &0.11 &0.06  \\
-38.81 &1.59 &1.24 &\nodata &\nodata &\nodata &\nodata &\nodata &\nodata &\nodata &\nodata \\
-37.04 &5.48 &4.26 &\nodata &\nodata &\nodata &\nodata &\nodata &\nodata &\nodata &\nodata \\
-33.52 &\nodata &\nodata &\nodata &\nodata &\nodata &\nodata &\nodata &\nodata &0.13 &0.08 \\
-22.93 &0.874 &0.680  &\nodata &\nodata &\nodata &\nodata &\nodata &\nodata &\nodata &\nodata \\
-17.64 &\nodata &\nodata &\nodata &\nodata &\nodata &\nodata &0.240 &\nodata &\nodata \\
-5.29  &\nodata &\nodata &0.0536 &0.0413 &0.271 &0.207 &\nodata&\nodata&0.22 &0.1319 \\
-3.53  &\nodata &\nodata &0.090 &0.070 &0.0551 &0.0421 &\nodata &\nodata &\nodata &\nodata \\
-1.76  &\nodata &\nodata &0.962 &0.741 &0.0023 &0.00177 &\nodata&\nodata&0.17 &0.10 \\
0.0   &\nodata &\nodata &0.0691 &0.0533 &0.113 &0.087 &\nodata &\nodata &\nodata &\nodata \\
1.76  &\nodata &\nodata &\nodata &\nodata &\nodata &\nodata &\nodata &\nodata &0.13 &0.08  \\
3.53  &\nodata &\nodata &\nodata &\nodata &\nodata &\nodata &0.55 &0.33  &0.3  &0.17  \\
5.29  &\nodata &\nodata &\nodata &\nodata &\nodata &\nodata &0.36 &0.218 &\nodata &\nodata \\
8.82  &\nodata &\nodata &0.583 &0.449 &0.87 &0.658 &1.59 &0.97 &2  &1.3 \\
10.58 &\nodata &\nodata &\nodata &\nodata &\nodata &\nodata &0.67 &0.41 &0.3 &0.20 \\
19.4  &\nodata &\nodata &\nodata &\nodata &\nodata &\nodata &\nodata &\nodata &0.18 &0.10 \\
21.17 &\nodata   &\nodata   &\nodata &\nodata &\nodata &\nodata &0.54 &0.33 &\nodata &\nodata \\
24.7  &1.63 &1.26 &0.159 &0.123 &1.15&0.88&\nodata&\nodata&0.18 &0.11 \\
26.46 &0.290 &0.226 &\nodata &\nodata &\nodata &\nodata &\nodata &\nodata &\nodata &\nodata \\
28.22 &0.0620 &0.0482 &\nodata &\nodata &\nodata &\nodata &\nodata &\nodata &0.15 &0.09 \\
29.99 &0.0938 &0.0730 &\nodata &\nodata &\nodata &\nodata &0.51 &0.311 &\nodata &\nodata\\
31.75 &3.82 &2.97 &0.118 &0.091 &0.291 &0.222 &\nodata&\nodata&0.09 &0.05 \\
33.52 &5.47 &4.25 &0.204 &0.157 &0.216 &0.165 &\nodata &\nodata &0.23 &0.13 \\
35.28 &\nodata &\nodata &0.096 &0.074 &0.237 &0.181 &\nodata &\nodata &\nodata &\nodata \\
38.81 &\nodata &\nodata &\nodata &\nodata &\nodata &\nodata &0.43&0.259 &\nodata &\nodata \\
42.34 &\nodata &\nodata &\nodata &\nodata &\nodata &\nodata &\nodata &\nodata &\nodata &\nodata \\
56.45 &\nodata &\nodata &0.091 &0.070 &0.230 &0.175 &\nodata &\nodata &\nodata &\nodata \\
58.21 &\nodata &\nodata &\nodata &\nodata &\nodata &\nodata &\nodata &\nodata &0.22 &0.13  \\
59.98 &\nodata &\nodata &0.077 &0.0587 &0.134 &0.102 &0.37 &0.226 &\nodata &\nodata \\
\enddata
\tablenotetext{a}{The negative values are for the northern regions of the spectra}
\tablenotetext{b}{$H_{\alpha}$ flux is corrected for corresponding $[NII]\lambda 6548$ flux/$H_{\beta}$}
\end{deluxetable}

\clearpage

\begin{table}
\caption{Oxygen and Sulphur abundance parameters and oxygen abundances for the polar disk in NGC4650A}   % title
                                                                                                         % of
                                                                                                         % Table
\label{rs23}   % is used to refer this table in the text
\centering
\begin{tabular}{l c c}        % centered columns (4 columns)
\hline\hline                 % inserts double horizontal lines
r (arcsec) & $R_{23}$ ($\pm 0.05$) & $12+log(O/H)_{P}$ ($\pm 0.2$)\\ % table heading
\hline                        % inserts single horizontal line
   & & \\
Slit 1 & & \\
       -44.1    &  10.05   &   7.8 \\
       -42.34    &  7.91   &   8.0 \\
       -40.57    &  7.70   &   8.1 \\
       -38.81    &  6.98   &   8.1 \\
       -37.04    &  7.35   &   8.1 \\
       -35.28    &   6.97   &    8.2 \\
       -33.52    &  7.04   &    8.2 \\
       -31.75    &  6.45   &    8.2 \\
       -29.99    & 6.57    &   8.2 \\
       -28.22    &  7.50   &    8.1 \\
       -17.64    &   9.33   &   8.0 \\
       -14.11    &  5.87   &   8.2 \\
       -3.528    &    8.90   &   8.0 \\
       37.04    &  11.33   &   8.0 \\
       38.81    &  13.64   &   7.8 \\
       45.86    &  3.86   &   8.4 \\
       47.63    &  6.20   &  8.0 \\
       49.39    &  5.34   &   8.1 \\
 & & \\
Slit 2 & & \\
       -33.52    &  3.18   &  8.5 \\
       -31.75    &  3.61   &  8.4 \\
       -3.528    &  6.97   &  8.3 \\
       5.292    &  8.08   &  8.1 \\
        8.82    &  5.93   &  8.2 \\
       17.64    &  7.10   &  8.2 \\
       21.17    &  6.24   &  8.4 \\
       22.93    &  6.28   &  8.2 \\
        24.7    &  6.73   &  8.3 \\
       26.46    &   7.26   &  8.2 \\
       28.22    &  8.30   &  8.0 \\
       29.99    &   6.81   &  8.2 \\
       31.75    &  10.74   &  7.7 \\
        44.1    &  6.96   &  8.3 \\
       45.86    &  5.91   &  8.1 \\
       47.63    &  5.55   &  8.4 \\
       56.45    &  9.13   &  8.1 \\
       58.21    &   10.30   & 8.0 \\
       59.98    &  14.48   &  7.7 \\
\hline                                   %inserts single line
\end{tabular}
\end{table}

\clearpage

\begin{table}
\caption{Oxygen abundances for the polar disk in NGC4650A derived by the $T_e$ method}   % title of Table
\label{OHT}   % is used to refer this table in the text
\centering
\begin{tabular}{l c c}        % centered columns (4 columns)
\hline\hline                 % inserts double horizontal lines
r (arcsec) & $12+log(O/H)_{T_{mod}}$ & $12+log(O/H)_{T_{obs}} (\pm 0.1)$\\ % table heading
\hline                        % inserts single horizontal line
   &   &  \\
 Slit 1 & & \\
       -44.10   &    $7.83  \pm    0.58$ & 8.6\\
       -42.34   &    $7.71  \pm    0.52$ & 8.5\\
       -40.57   &    $7.69  \pm    0.49$ & 8.5\\
       -38.81   &    $7.65  \pm    0.50$ & 8.4\\
       -37.04   &    $7.67  \pm    0.46$ & 8.4\\
       -35.28   &    $7.64  \pm    0.46$ & 8.4\\
       -33.52   &    $7.65  \pm    0.45$ & 8.4\\
       -31.75   &    $7.62  \pm    0.49$ & 8.4\\
       -29.99   &    $7.63  \pm    0.49$ & 8.4\\
       -28.22   &    $7.69  \pm    0.50$ & 8.5\\
       -17.64   &    $7.76  \pm    0.42$ & 8.5\\
       -14.11   &    $7.58  \pm    0.49$ & 8.4\\
       -3.53   &    $7.75  \pm    0.46$ & 8.5\\
       37.04   &    $7.80  \pm    0.33$ & 8.4\\
       38.81   &    $7.88  \pm    0.33$ & 8.5\\
       45.86   &    $7.42  \pm    0.58$ & 8.2\\
       47.63   &    $7.64  \pm    0.68$ & 8.5\\
       49.39   &    $7.58  \pm    0.67$ & 8.4\\
  & &\\
 Slit 2 & &\\
       -33.52   &    $7.34  \pm    0.57$ & 8.1\\
       -31.75   &    $7.39  \pm    0.57$ & 8.2\\
        -3.53   &    $7.61  \pm    0.36$ & 8.3\\
        5.29   &    $7.70  \pm    0.42$ & 8.4\\
        8.82   &    $7.59  \pm    0.52$ & 8.4\\
       17.64   &    $7.64  \pm    0.42$ & 8.4\\
       21.17   &    $7.56  \pm    0.35$ & 8.2\\
       22.93   &    $7.61  \pm    0.48$ & 8.4\\
       24.70   &    $7.59  \pm    0.36$ & 8.3\\
       26.46   &    $7.64  \pm    0.40$ & 8.4\\
       28.22   &    $7.72  \pm    0.47$ & 8.5\\
       29.99   &    $7.63  \pm    0.46$ & 8.4\\
       31.75   &    $7.85  \pm    0.54$ & 8.6\\
       44.10   &    $7.61  \pm    0.36$ & 8.3\\
       45.86   &    $7.62  \pm    0.68$ & 8.5\\
       47.63   &    $7.51  \pm    0.35$ & 8.2\\
       56.45   &    $7.73  \pm    0.37$ & 8.4\\
       58.21   &    $7.78  \pm    0.38$ & 8.5\\
       59.98   &    $7.92  \pm    0.36$ & 8.6\\

\hline                                   %inserts single line
\end{tabular}
\end{table}

%% the file, uncomment the \plotone, \plottwo, and
%% If you need a layout that cannot be achieved with \plotone or
%% \plottwo, you can invoke the graphicx package directly with the
%% \includegraphics command or use \plotfiddle. For more information,
%% please see the tutorial on "Using Electronic Art with AASTeX" in the
%% documentation section at the AASTeX Web site,
%% http://www.journals.uchicago.edu/AAS/AASTeX.
%%
%% The examples below also include sample markup for submission of
%% supplemental electronic materials. As always, be sure to check
%% the instructions to authors for the journal you are submitting to
%% for specific submissions guidelines as they vary from

%% This example uses \plotone to include an EPS file scaled to
%% 80% of its natural size with \epsscale. Its caption
%% has been written to indicate that additional figure parts will be
%% available in the electronic journal.

%\begin{figure*}
%\centering
%\includegraphics[width=10cm]{ratio.ps}
%\caption{i triangoli sono il sud e i quadrati sono il nord}
%\label{ROIII}
%\end{figure*}

%\begin{figure*}
%\centering
%\includegraphics[width=10cm]{OIII_all.ps}
%\caption{...} \label{temp}
%\end{figure*}

%\begin{figure*}
%\centering
%\includegraphics[width=10cm]{OH_T.ps}
%\caption{Abbondanza di ossigeno calcolata usando la stima della
%temperatura in \citet{Hag08}} \label{OH_T}
%\end{figure*}

%% Here we use \plottwo to present two versions of the same figure,
%% one in black and white for print the other in RGB color
%% for online presentation. Note that the caption indicates
%% that a color version of the figure will be available online.
%%

%% If you are not including electonic art with your submission, you may
%% mark up your captions using the \figcaption command. See the
%% No more than seven \figcaption commands are allowed per page,
%% so if you have more than seven captions, insert a \clearpage

%% Tables should be submitted one per page, so put a \clearpage before

%% Two options are available to the author for producing tables:  the
%% deluxetable environment provided by the AASTeX package or the LaTeX
%% table environment.  Use of deluxetable is preferred.
%%

%% Three table samples follow, two marked up in the deluxetable environment,
%% one marked up as a LaTeX table.

%% In this first example, note that the \tabletypesize{}
%% command has been used to reduce the font size of the table.
%% We also use the \rotate command to rotate the table to
%% landscape orientation since it is very wide even at the
%% Note also that the \label command needs to be placed

%% This table also includes a table comment indicating that the full
%% version will be available in machine-readable format in the electronic

\clearpage

%% If you use the table environment, please indicate horizontal rules using
%% Do not put multiple tabular environments within a single table.
%% The optional \label should appear inside the \caption command.

\clearpage

%% Any table notes must follow the \end{tabular} command.

%% If the table is more than one page long, the width of the table can vary
%% from page to page when the default \tablewidth is used, as below.  The
%% individual table widths for each page will be written to the log file; a
%% maximum tablewidth for the table can be computed from these values.
%% The \tablewidth argument can then be reset and the file reprocessed, so
%% that the table is of uniform width throughout. Try getting the widths
%% from the log file and changing the \tablewidth parameter to see how
%% adjusting this value affects table formatting.

%% The \dataset{} macro has also been applied to a few of the objects to
%% show how many observations can be tagged in a table.

\clearpage

%% Tables may also be prepared as separate files. See the accompanying
%% sample file table.tex for an example of an external table file.
%% To include an external file in your main document, use the \input
%% command. Uncomment the line below to include table.tex in this
%% sample file. (Note that you will need to comment out the \documentclass,
%% \begin{document}, and \end{document} commands from table.tex if you want
%% to include it in this document.)


%% The following command ends your manuscript. LaTeX will ignore any text

\end{document}